\documentclass[12pt]{article}
\usepackage{graphicx}
\newcommand{\fslash}[1]{\mbox{$\!\not\!#1$}}

\newcommand{\bold}[1]{\mbox{\boldmath ${#1}$}}

\begin{document}

\baselineskip 4 ex

\title{Medium modifications of nucleon electromagnetic form factors}
       
\author{T. Horikawa  \\
       Department of Physics, School of Science, Tokai University \\
       Hiratsuka-shi, Kanagawa 259-1292, Japan \\      
{ } \\
       W. Bentz \thanks{Correspondence to: W. Bentz, E-mail: 
bentz@keyaki.cc.u-tokai.ac.jp}\\
       Department of Physics, School of Science, Tokai University \\
       Hiratsuka-shi, Kanagawa 259-1292, Japan}

\date{ }
\maketitle
\begin{abstract}
We use the Nambu-Jona-Lasinio model as an effective quark theory
to investigate the medium modifications of the nucleon electromagnetic
form factors. By using the equation of state of nuclear matter
derived in this model, we discuss the results based on the naive 
quark - scalar diquark picture, the effects of finite diquark size, 
and the meson cloud around the constituent quarks. We apply this 
description to the longitudinal response function
for quasielastic electron scattering. RPA correlations, based on
the nucleon-nucleon interaction derived in the same model, are also
taken into account in the calculation of the response function.

\vspace{0.3 cm}

\noindent
{\footnotesize PACS numbers: 12.39.Fe; 12.39.Ki; 13.40.Gp; 14.20.Dh; 
14.65.-q }
\end{abstract}

\newpage

\section{Introduction}
\setcounter{equation}{0}

The structure of the nucleon and its modifications in the nuclear
medium is a very active field of experimental and theoretical
research. The basic quantities, which reflect the charge and
current distributions in the nucleon, are the electromagnetic
form factors \cite{TW}, which are currently investigated in elastic electron-
nucleon scattering experiments from intermediate to very high energies
\cite{FORMEX}.
The knowledge of the nucleon form factors is also inevitable to
understand the electromagnetic structure of nuclei. Electron-nucleus 
scattering experiments under quasielastic kinematic 
conditions, like the measurement of inclusive response functions
in the intermediate energy region \cite{MEZ,WILL} and recent measurements of
polarization transfer in semi-exclusive knock-out processes \cite{POL}, are
ideal places to study the form factors of a nucleon bound in the
nuclear medium. Because the structure of the quark core and the
surrounding meson cloud may be different for a bound nucleon
and a free nucleon, one expects medium modifications of the
nucleon form factors\cite{MEDF}, and the exploration of these effects is
an important subject at current electron accelerator facilities
\cite{JLAB}.

On the theoretical side, effective quark theories are the ideal
tools to describe the electromagnetic form factors of the nucleon.
Much progress for the case of the free nucleon has been made in 
Faddeev type descriptions based on the Schwinger-Dyson method \cite{SD}.
An important point which still has to be implemented in these
calculations is the role of the pion cloud around the nucleon,
and the recently developed method of chiral extrapolations
of lattice results \cite{CHIR} provides important hints. On the other hand,
the calculation of form factors at finite nucleon density requires
also a description of the equation of state of the many-nucleon system, 
and here progress has been made by using the Nambu-Jona-Lasinio (NJL)
model \cite{NJL} as an effective quark theory: Recent works have shown how
to account for the saturation properties of nuclear matter in
this model \cite{BT}, and when combined with the quark-diquark description of
the single nucleon\cite{MBY} this provides a successful description of both
nucleon and nuclear structure functions for deep inelastic scattering
\cite{EMC,NEWI}
\footnote{Recently the model has been extended to describe the equation of
state at high densities\cite{QM,NEWS}.}. 

The purpose of this paper is to discuss the results for the nucleon 
form factors obtained in the simple quark-scalar diquark description
of the nucleon at finite density in the NJL model. We have to note
from the beginning that this can only be a first step toward a
realistic description, because it is known that axial vector
diquarks are important for spin-dependent quantities\cite{AX,NEWI}, 
and the pion cloud is important for magnetic moments
and the size of the nucleon\cite{CHIR,BAB}. While the axial diquarks 
could be included in a further step like it was done for the structure
functions\cite{NEWI}, a reliable description of pion cloud effects makes it
necessary to go beyond the standard ladder approximation scheme.
However, like the simple quark-scalar diquark model of Ref.\cite{EMC}
served as a
basis for the more elaborate description of structure functions
\cite{NEWI}, it will also be the basis of a more realistic description 
of form factors
including axial diquarks and the pion cloud. To provide this basis 
is the main intention of the present paper.

In Sect. 2 we will briefly review the model for the nucleon and the
nuclear matter equation of state. Sect. 3 is devoted to the
nucleon form factors at finite density, and in Sect.4 we discuss the
numerical results. As an application, we
discuss the response function for quasielastic electron scattering
in Sect.5. For this purpose we will also elucidate the nucleon-nucleon
interaction in our model in order to include the correlations within
the relativistic RPA. A summary will be presented in Sect. 6.   

\section{The model}
\setcounter{equation}{0}
In this work, we use the NJL model as an effective quark
theory to describe the nucleon as a quark-diquark bound state, and 
nuclear matter (NM) in the
mean field approximation. The details are explained in Refs. \cite{BT,QM},
and here we will only briefly summarize those points which will be
needed for our calculations.

The NJL model is characterized by a chirally symmetric 4-fermi 
interaction between the quarks\cite{NJL1}. Any such interaction 
can be Fierz symmetrized and decomposed into various
$q\overline{q}$ channels \cite{FAD}. Writing out explicitly only those 
channels which are relevant for our present discussion, we have
\begin{equation}
{\cal L}={\overline \psi}\left(i \fslash{\partial}-m\right) \psi + G_{\pi}
\left( \left(\overline{\psi}\psi \right)^2
	- \left(\overline{\psi}(\gamma_5\bold{\tau})\psi\right)^2 \right)
- G_{\omega} \left( \overline{\psi} \gamma^{\mu} \psi \right)^2  + \ldots
\label{lag}
\end{equation}
where $m$ is the current quark mass.
In a mean field description of the isospin symmetric nuclear matter 
ground state $|\rho \rangle$, the Lagrangian can be expressed as 
\begin{eqnarray}
{\cal L}={\overline \psi}\left(i \fslash{\partial}- M - \fslash{V}
\right) \psi - \frac{(M-m)^2}{4 G_{\pi}} + \frac{V_{\mu} V^{\mu}}{4G_{\omega}} + {\cal L}_I,
\label{lag1}
\end{eqnarray}
where $M=m-2G_{\pi} \langle \rho|\overline{\psi}\psi|\rho \rangle$ and
$V^{\mu}= 2 G_{\omega} \langle \rho|\overline{\psi}\gamma^{\mu}\psi|\rho 
\rangle$, and ${\cal L}_I$ is the normal ordered interaction Lagrangian.
The effect of the mean scalar field is thus included in the 
density-dependent constituent quark mass $M$, and the effect of the
mean vector field is to shift the quark momentum according to
$p^{\mu} = p_{Q}^{\mu} + V^{\mu}$, where $p_{Q}^{\mu}$ is the kinetic
momentum. The propagator of the constituent quark therefore has the following
dependence on the mean vector field
\footnote{In this section, Green functions
in the presence of the mean vector field are denoted by a tilde, and
those without the vector field have no tilde. In the loop integrals for 
the electromagnetic form factors in Sect.3, however, it is always possible 
to eliminate the vector field by a shift of the integration variable,
and therefore the tilde-Green functions do not appear in later sections.}
: $\tilde{S}(k) = S(k_{\rm Q})$.

One can use a further Fierz transformation to
decompose ${\cal L}_I$ into a sum of
$qq$ channel interaction terms \cite{FAD}. For
our purposes we need only the interaction in the scalar diquark
($J^{\pi}=0^+, T=0$, color ${\overline 3}$) channel:
\begin{eqnarray}
{\cal L}_{I,s} = G_s \left(\overline{\psi}\left(\gamma_5 C\right)\tau_2
\beta^A \overline{\psi}^T\right) \left(\psi^T\left(C^{-1}\gamma_5\right)
\tau_2 \beta^A \psi\right),
\label{lags}
\end{eqnarray}
where $\beta^A=\sqrt{3/2}\,\, \lambda^A \,\,(A=2,5,7)$ are the color
${\overline 3}$ matrices and $C=i\gamma_2 \gamma_0$.
The coupling constant $G_s$ will be
determined so as to reproduce the free nucleon mass.

The reduced t-matrix in the scalar diquark channel is given by \cite{BT}
\begin{eqnarray}
\tilde{\tau}_s(q)=\frac{4iG_s}{1 + 2 G_s \tilde{\Pi}_{s}(q)} = 
\tau_s(q_{\rm D})
\label{taus}
\end{eqnarray}
with the scalar $qq$ bubble graph
\begin{eqnarray}
\tilde{\Pi}_{s}(q)=6 i \int \frac{{\rm d}^4 k}{(2\pi)^4} {\rm tr}_D \left[
	\gamma_5 S(k) \gamma_5  S \left(-(q-k)\right) \right]
= \Pi_{s}(q_{\rm D})\,.
\label{bubbs}
\end{eqnarray}
Here $q_{\rm D}^{\mu}=q^{\mu}-2V^{\mu}$ is the kinetic momentum 
of the diquark. 

The relativistic Faddeev equation in the NJL model can been solved numerically
for the free nucleon\cite{FAD}, but here we restrict ourselves to the 
static approximation\cite{STAT}, where the momentum dependence of the 
quark exchange kernel is neglected. The solution for the quark-diquark 
t-matrix then takes the simple analytic form
\begin{eqnarray}
\tilde{T}_N(p)=\frac{3}{M} \frac{1}{1+\frac{3}{M} \tilde{\Pi}_{N}(p)} = 
T_N(p_{\rm N})\,,
\label{tn}
\end{eqnarray}
with the quark-diquark bubble graph given by
\begin{eqnarray}
\tilde{\Pi}_{N}(p)=-\int \frac{{\rm d}^4 k}{(2\pi)^4}\, \tilde{S}(k)\, 
\tilde{\tau}_s(p-k) = \Pi_{N}(p_{\rm N})\,,
\label{bubbn}
\end{eqnarray}
where $p_{\rm N}^{\mu}=p^{\mu}-3V^{\mu}$ is the kinetic momentum 
of the nucleon. The nucleon mass $M_{N}$ is defined as the
pole of (\ref{tn}) at $\fslash{p}_{\rm N}=M_N$,
and the positive energy spectrum has the form $p_0 = \epsilon_p \equiv 
E_{Np} + 3 V_0$ with $E_{Np} = \sqrt{M_N^2 + {\bold p}^2}$. 
The nucleon vertex function in the non-covariant normalization
is defined by the pole behavior of the quark-diquark t-matrix:
\begin{eqnarray}
T_N(p) \rightarrow \frac{\Gamma_N(p) \, \overline{\Gamma}_N(p)}
{p_0 - \epsilon_p} \,\,\,\,\,\,\,\,\,\,\,\,\,\,\,{\rm as}\,\,\,
p_0 \rightarrow \epsilon_p      \label{lcn}\,.
\end{eqnarray}
From this definition and Eq.(\ref{tn}) one obtains
\begin{eqnarray}
\Gamma_N(p)=\sqrt{- Z_N \, \frac{M_N}{E_{Np}}} \,\, u_N(p_{\rm N})\,,  
\label{gamma}
\end{eqnarray}
where $u_N$ is a free Dirac spinor for mass $M_N$ normalized as 
$\overline{u}_N u_N = 1$. The normalization factor $Z_N$ is easily obtained
from this definition and will be given in Eq.(\ref{zn}) below.
We note that with this normalization the vertex function satisfies the relation
\begin{eqnarray}
\overline{\Gamma}_N(p) \, \left(\frac{\partial \Pi_{N}(p)}
{\partial p_{\mu}}\right) \, \Gamma_N(p) = \frac{p^{\mu}}{E_{Np}} \label{nwf}
\end{eqnarray}
In the numerical calculations of this paper, we will approximate the 
quantity $\tau_s$ by a ``contact+pole'' form:
\begin{eqnarray}
\tau_s(q) \rightarrow 4i G_s - \frac{i\,g_s}{q^2-M_s^2} \equiv
4i G_s -i g_s\, \Delta_{Fs}(q). \label{pole}
\end{eqnarray}
Here $\Delta_{Fs}(q)$ is the Feynman propagator for a scalar particle
of mass $M_s$, which is defined as the pole of $\tau_s$ of 
Eq.(\ref{taus}). The residue at the pole ($g_s$) will be given in Eq.
(\ref{gs}) below. 

In the calculation of the nucleon form factors, we will also consider the
effects of the pion cloud around the constituent quarks. In this case,
the propagator $S(p)$ of the quark  involves an additional self energy 
correction from the pion cloud ($\Sigma_Q$). Here we will use a 
simple pole approximation for $S(p)$: 
\begin{eqnarray}
S(p) = Z_Q\,S_F(p)\,, \,\,\,\,\,\,\,\,\,\,
Z_Q^{-1} = 1 - \left(\frac{\partial \Sigma_Q}{\partial \fslash{p}}
\right)_{\fslash{p}=M}\,,  \label{self}
\end{eqnarray}
where $S_F$ is the Feynman propagator of a constituent quark with mass $M$.
In this approximation, the pion effects can be renormalized by 
$\psi\rightarrow \sqrt{Z_Q} \psi$ and a redefinition of the four fermi
coupling constants $G_{\alpha} \rightarrow G_{\alpha}/Z_Q^2$, 
see Ref.(\cite{MBY}). For the calculation of the form factors, however,
we will keep the factor $Z_Q$ explicitly for clarity
\footnote{We also note that such a renormalization procedure
is no longer possible when one considers the pion cloud effects
around the {\em nucleon}, which goes beyond the simple ladder approximation
on the quark level.}.  
Introducing (\ref{self}) and (\ref{pole}) into the expressions 
(\ref{bubbs}) and (\ref{bubbn}) for the bubble graphs shows that the 
diquark and nucleon normalization factors can be written as
\begin{eqnarray}
g_s^{-1} &=& \frac{1}{2} \, \left( - \frac{\partial \Pi_{s}(q)}{\partial q^2}
\right)_{q^2=M_s^2} =  \frac{1}{2} \, Z_Q^2 \, 
\left( - \frac{\partial \hat{\Pi}_{s}(q)}{\partial q^2}
\right)_{q^2=M_s^2} \equiv  \, Z_Q^2 \, \hat{g}_s^{-1}
\nonumber \\  \label{gs} \\
Z_N^{-1} &=& \left(- \frac{\partial \Pi_{N}(p)}{\partial \fslash{p}}\right)
_{\fslash{p}=M_N} = Z_Q\,g_s \,
\left(- \frac{\partial \hat{\Pi}_{N}(p)}{\partial \fslash{p}}\right)
_{\fslash{p}=M_N} \equiv Z_Q\,g_s \, \hat{Z}_N^{-1}\,, \nonumber \\
\label{zn}
\end{eqnarray}
where $\hat{\Pi}_s$ and $\hat{\Pi}_N$ are defined
in terms of the pole parts only:
\begin{eqnarray}
\hat{\Pi}_{s}(q) &=& 6 i \int \frac{{\rm d}^4 k}{(2\pi)^4} {\rm tr}_D \left[
	\gamma_5 S_F(k) \gamma_5  S_F \left(-(q-k)\right) \right]
\label{newbubbs} \\
\hat{\Pi}_{N}(p)&=&i \int \frac{{\rm d}^4 k}{(2\pi)^4}\, S_F(k)\, 
\Delta_{Fs}(p-k)  \label{newbubbn} 
\end{eqnarray}

The equation of state of NM in the NJL model can be derived in a formal way
\cite{QM} from the quark Lagrangian (\ref{lag1}) by using hadronization
techniques, but in the mean field approximation the resulting energy
density of isospin symmetric NM has the simple form \cite{BT}
\begin{eqnarray}
{\cal E}={\cal E}_V - \frac{V_{0}^2}{4G_{\omega}} + 4 \int
\frac{d^3 p}{(2 \pi)^3}\, \Theta\left(p_F-|{\bold p}|\right)\, \epsilon_p,
\label{en}
\end{eqnarray}
where the vacuum contribution (quark loop) is
\begin{eqnarray}
{\cal E}_V=12i \int \frac{{\rm d}^4 k}{(2\pi)^4} \,
{\rm ln}\, \, \frac{k^2-M^2+i\epsilon}{k^2-M_0^2+i\epsilon}
+\frac{(M-m)^2}{4 G_{\pi}} -\frac{(M_0-m)^2}{4 G_{\pi}}\,.
\label{env}
\end{eqnarray}
Here $M_0$ the constituent quark mass for zero nucleon density.
The condition $\partial {\cal E} / \partial V_0 = 0$ leads to
$V_0=6 \, G_{\omega}\, \rho$, and we can eliminate the vector field 
in (\ref{en})
in favor of the baryon density. The resulting expression has then
to be minimized with respect to the constituent quark
mass $M$ for fixed density. For zero density this
condition becomes identical to the
familiar gap equation of the NJL model\cite{NJL}, and for finite
density the nonlinear $M$-dependence of the
nucleon mass $M_N$ is
essential to obtain saturation of the binding energy per nucleon~\cite{BT}.

In order to fully define the model one has to specify a cut-off procedure.
In the calculations in this paper
we will use the proper time regularization
scheme~\cite{IR,BT}, where one evaluates
loop integrals over a product of propagators by
introducing Feynman parameters, performing a Wick rotation and
replacing the denominator ($A$) of the loop integral according to
\begin{eqnarray}
\frac{1}{A^n} \rightarrow \frac{1}{(n-1)!} 
\int_{1/\Lambda_{\rm UV}^2}^{1/\Lambda_{\rm IR}^2}
{\rm d}\tau \, \tau^{n-1}\,e^{-\tau A}\,\,\,\,\,\,\,\,\,(n\geq 1), \label{pt}
\end{eqnarray}
where $\Lambda_{\rm IR}$ and $\Lambda_{\rm UV}$ are the infrared 
and  ultraviolet cut-offs, respectively.
The infrared cut-off plays the important role of
eliminating the unphysical thresholds for the decay of the 
nucleon and mesons into quarks \cite{IR}, thereby taking into 
consideration a particular aspect of confinement physics.

\section{Nucleon electromagnetic form factors}
\setcounter{equation}{0}

The electromagnetic current of the nucleon in the quark-diquark model
is represented by the Feynman diagrams of Fig.1 and given by
\begin{eqnarray}
j_N^{\mu}(q) = \overline{\Gamma}_N(p') \int \frac{{\rm d}^4 k}{(2\pi)^4}
\left[ \left(S(p'-k) \Lambda_Q^{\mu} S(p-k)\right) \tau_s(k) \right.
\nonumber \\ 
\left. + i \left(\tau_s(p'-k) \Lambda_D^{\mu} \tau_s(p-k)\right) S(k) \right] 
\Gamma_N(p)\,.
\label{curr}
\end{eqnarray}

\begin{figure}[ht]
\begin{center}
\includegraphics[scale=0.9]{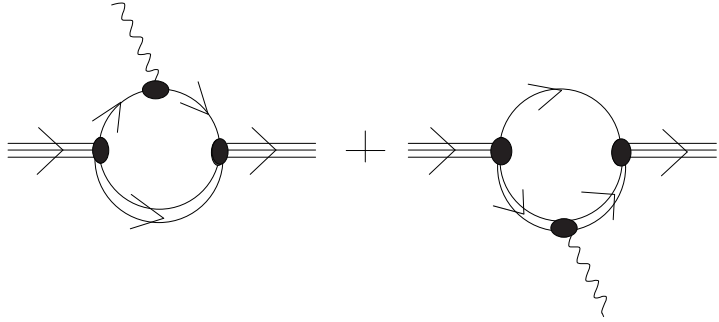}
\caption{Feynman graphs for the evaluation of the 
nucleon electromagnetic
current in the quark-diquark model. The double line represents the diquark
t-matrix, the solid line the constituent quark propagator, and the black
areas are the vertex functions and electromagnetic vertices.}
\end{center}
\end{figure}

Here $\Lambda_Q^{\mu}$ and $\Lambda_D^{\mu}$ are the electromagnetic vertices
of the quark and the scalar diquark, both depending on the final
and initial particle momenta. It is an easy task to use the Ward-Takahashi
identities \cite{WARD} for the quark and diquark vertices
\begin{eqnarray}
q_{\mu} \left(S(\ell') \Lambda_Q^{\mu} S(\ell)\right) &=& 
- Q_Q \left(S(\ell') - S(\ell)\right) \label{wtq}  \\
q_{\mu} \left(\tau_s(\ell') \Lambda_D^{\mu} \tau_s(\ell)\right) &=& i Q_D
\left(\tau_s(\ell') - \tau_s(\ell)\right) \label{wtd}
\end{eqnarray}
to show the current and charge conservation for the nucleon:
\begin{eqnarray}
q_{\mu} j_N^{\mu}(q) &=& Q_N \overline{\Gamma}_N(p') \left(\Pi_N(p') -
\Pi_N(p) \right) \Gamma_N(p) = 0 \label{cc}  \\
j^{\mu}(0) &=& Q_N \, \frac{p^{\mu}}{E_{Np}}\,, \label{chc} 
\end{eqnarray}
where the electric charge of the nucleon $Q_N=Q_Q + Q_D$ is the sum of
the quark and diquark electric charges. 

The electromagnetic vertices in (\ref{curr}) describe
the finite extension of the constituent quarks and the diquark. In 
general, they should be calculated off-shell consistently 
with the propagators $\tau_s$ and $S$, using Feynman diagrams or some ansatz 
which satisfies the Ward identities\cite{SD}. In the present work,
however, our principal aim is to investigate the medium effects in a
simple model calculation. For this purpose, we limit the complications
caused by the quark and diquark sizes to a minimum, and 
use an on-shell (or pole) approximation for the quark and diquark
currents appearing in Eq.(\ref{curr}): 
\begin{eqnarray}
\left(S(\ell') \, \Lambda_Q^{\mu} \, S(\ell)\right) &\longrightarrow&
Z_Q \left(S_F(\ell')\, \hat{\Lambda}_{Q}^{\mu}\, S_F(\ell)\right) 
\label{poleq} \\
\left(\tau_s(\ell')\, \Lambda^{\mu}_D \, \tau_s(\ell)\right) &\longrightarrow&
- g_s \left(\Delta_{Fs}(\ell') \, \hat{\Lambda}_{D}^{\mu} \, 
\Delta_{Fs}(\ell) 
\right)\,,
\label{poled}
\end{eqnarray}
where the on-shell (o.s.) vertices are denoted by a hat and given by the 
pole residues of the full quantities by
\begin{eqnarray}
\hat{\Lambda}_{Q}^{\mu} &\equiv& Z_Q \left(\Lambda_{Q}^{\mu}\right)_
{\rm o.s.} \equiv
\gamma^{\mu}\, F_{1Q}(q^2) + \frac{ i \sigma^{\mu \nu}q_{\nu}}{2M} \,
F_{2Q}(q^2)
\label{quarkff} \\   
\hat{\Lambda}_{D}^{\mu} &\equiv& g_s \left(\Lambda_{D}^{\mu}\right)_
{\rm o.s.}
\equiv (\ell'+\ell)^{\mu} \, F_D(q^2). \label{diquarkff}
\end{eqnarray}
Here we introduced the quark and diquark form factors which satisfy 
$F_{1Q}(0) = Q_Q, \,\,F_D(0) = Q_D$. 

To understand (\ref{poleq}) and (\ref{poled}), we note that
in general the on-shell approximation for a vertex function 
$\Lambda^{\mu}(\ell',\ell)$ can be formulated
only if it appears between pole parts of Green functions, because only in
this case one can approximate it by its value for
on-shell momenta $\ell',\,\ell$ 
\footnote{This approximation is the basis of the standard  
convolution formalism
to calculate quark light-cone momentum distributions and structure 
functions.}. 
This is why in Eq.(\ref{poleq}) and (\ref{poled}) we have
replaced the propagators left and right to the vertex functions by their 
pole parts (see (\ref{pole}) and (\ref{self})), which is
also essential in order to have charge conservation with
the vertices (\ref{quarkff}) and (\ref{diquarkff}). 

We now can write down the form of the nucleon
current (\ref{curr}) which will be used in the further calculations: 
\begin{eqnarray}
j_N^{\mu}(q) = \sqrt{\frac{M_N}{E_{Np}}\,\frac{M_N}{E_{Np'}}} 
\, \overline{u}_N(p') \left({\cal O}_C^{\mu} + {\cal O}_Q^{\mu} + 
{\cal O}_D^{\mu} \right) \, u_N(p) \,.  \label{split}
\end{eqnarray}
Here the first and second terms denote the contributions of the contact term
and the pole term of the diquark t-matrix to the quark diagram
(first diagram of Fig.1), and the third term is the contribution 
from the diquark diagram:
\begin{eqnarray}
{\cal O}_C^{\mu} &=& - 
\frac{4i G_s}{\hat{g}_s} \hat{Z}_N \int \frac{{\rm d}^4 k}{(2\pi)^4}
S_F(p'-k) \left(\gamma^{\mu} F_{1Q}(q^2) + 
\frac{ i \sigma^{\mu \nu}q_{\nu}}{2M} F_{2Q}(q^2) \right) S_F(p-k) 
\nonumber \\
\label{contact} \\
{\cal O}_Q^{\mu} &=& i \hat{Z}_N \int \frac{{\rm d}^4 k}{(2\pi)^4}
S_F(p'-k) \left(\gamma^{\mu} F_{1Q}(q^2) + 
\frac{ i \sigma^{\mu \nu}q_{\nu}}{2M} F_{2Q}(q^2) \right) S_F(p-k) 
\Delta_{Fs}(k)
\nonumber \\
\label{quark} \\  
{\cal O}_D^{\mu} &=&  i \hat{Z}_N F_D(q^2) \int \frac{{\rm d}^4 k}{(2\pi)^4}
\Delta_{Fs}(p'-k) \left(p+p'-2k\right)^{\mu} \Delta_{Fs}(p-k)\,
S_F(k) \,. 
\label{diquark}
\end{eqnarray}
(For the contact term, we replaced $G_s \rightarrow G_s/Z_Q^2$, so that 
$G_s$ in (\ref{contact}) is the renormalized coupling in the sense explained
in Sect.2.)
By using the elementary Ward-Takahashi identities
$\fslash{q}=S_F^{-1}(\ell')-S_F^{-1}(\ell)$ and 
$\ell'^2 - \ell^2 = \Delta_{Fs}^{-1}(\ell') - \Delta_{Fs}^{-1}(\ell)$ and 
the fact that on the nucleon mass shell 
$\Pi_N(p)=\Pi_N(p')$, it is easy to check that the 3 parts in (\ref{split}) 
satisfy current conservation separately\footnote{These formal manipulations
involve shifts of the integration variables. In the actual calculations  
based on our regularization scheme, 
however, we always checked that current and charge conservation are
satisfied exactly. (Therefore, the explicit expressions given in the 
Appendices all satisfy the charge conservation.)}. 
Similarly, charge conservation can be checked by
using the elementary Ward identities
$\gamma^{\mu} = {\partial}S_F^{-1}(\ell)/\partial \ell^{\mu}$
and $2 \ell^{\mu} = \partial \Delta_{Fs}^{-1}(\ell)/\partial \ell^{\mu}$,
as well as $F_{1Q}(0)=Q_Q$, $F_D(0)=Q_D$. It has to be noted, however,
the general Ward-Takahashi identity for an {\em off-shell} nucleon
(the first equality in Eq.(\ref{cc}) without the nucleon spinor) 
is not valid in this approximation scheme.  

It is not very difficult to evaluate the three loop integrals in 
(\ref{contact})-(\ref{diquark}), and in Appendix A the results are given 
in terms
of the Dirac-Pauli form factors $F_{1N}$ and $F_{2N}$, which are defined
by 
\begin{eqnarray}
j_N^{\mu} &=& \sqrt{\frac{M_N}{E_{Np}}\,\frac{M_N}{E_{Np'}}} 
\, \overline{u}_N(p') \left[\gamma^{\mu} F_{1N}(q^2) + 
\frac{i \sigma^{\mu \nu}q_{\nu}}{2M_N} F_{2N}(q^2) \right] u_N(p)\,. 
\nonumber \\ 
\label{par1} 
\end{eqnarray}

In the following we will discuss various steps in the calculation of
the nucleon form factors.

\subsection{Naive quark-diquark model}
The simplest approximation consists in assuming point couplings of the 
quarks and diquarks to the photon, i.e., replacing 
${\displaystyle F_{1Q}\rightarrow Q_Q = \frac{1}{6}
+ \frac{\tau_3}{2}}$, ${\displaystyle F_{2Q} \rightarrow 0}$, 
${\displaystyle F_D \rightarrow Q_D = 
\frac{1}{3}}$ in Eqs. (\ref{contact})-(\ref{diquark}). This approximation
will be called the ``naive quark-diquark model'', and the detailed
expressions can be found in Appendix A.

\subsection{Effects of finite diquark size}

\begin{figure}[ht]
\begin{center}
\includegraphics[scale=0.6]{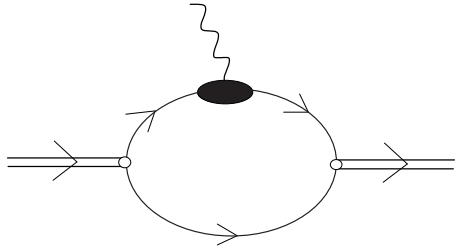}
\caption{Graphical representation of the diquark form factor. The white
circles denote the diquark vertex functions, and the black area is the
quark electromagnetic vertex. (There is a second diagram where the photon
couples to the other quark, but also an overall symmetry factor 
$\frac{1}{2}$.)}
\end{center}
\end{figure}

Here we consider the effect of the diquark form factor $F_D$, which 
has been defined in (\ref{diquarkff}). The
vertex $\Lambda^{\mu}_D$ is shown graphically in Fig.2, where the quark-diquark
vertex functions are those appearing in the Lagrangian(\ref{lags}).
We obtain
\begin{eqnarray}
\hat{\Lambda}_D^{\mu} = i \, g_s \int \frac{{\rm d}^4 k}{(2\pi)^4}
\left\{ {\rm Tr} \left[ \gamma_5 S(p'+k) \Lambda_Q^{\mu} S(p+k) \gamma_5
S(k) \right] \right\}\,, \label{df1}
\end{eqnarray}
where the trace refers to color, isospin and Dirac indices.
Using the on-shell approximation (\ref{poleq}), the definition
of quark form factors (\ref{quarkff}), and also (\ref{self}) and 
(\ref{gs}), we obtain
\begin{eqnarray}
\hat{\Lambda}_D^{\mu} &=& 6i \, {\hat g}_s \int \frac{{\rm d}^4 k}{(2\pi)^4}
\left\{ F_{1Q}^{(0)}(Q^2) {\rm Tr}_D \left[ \gamma_5 S_F(p'+k) \gamma^{\mu}
S_F(p+k) \gamma_5 S_F(k) \right] \right. \nonumber \\    
&+& \left. F_{2Q}^{(0)}(Q^2) {\rm Tr}_D \left[\gamma_5 S_F(p'+k) 
\frac{i \sigma^{\mu \nu} q_{\nu}}{2M} S_F(p+k) \gamma_5 S_F(k) \right]
\right\}\,.  \label{vert} 
\end{eqnarray}
Here $F_{1Q}^{(0)}$ and $F_{2Q}^{(0)}$ are the isoscalar parts of the
quark form factors $F_{1Q}$ and $F_{2Q}$. 
The resulting diquark form factor $F_D$ is given in Appendix A. 

\subsection{Effects of meson cloud around constituent quarks}
Here we consider the quark form factors arising from the pion cloud
and vector mesons. For the pion cloud, we obtain from the
definition (\ref{quarkff}) and the Feynman diagrams of Fig.3,
\begin{eqnarray}
\hat{\Lambda}_Q^{\mu} &=& Z_Q \gamma^{\mu} Q_Q
+ Z_Q \int \frac{{\rm d}^4 k}{(2\pi)^4} \left[
- \gamma_5 \tau_i \left(S(p'-k) \Lambda_{Q0}^{\mu}S(p-k)\right) 
\tau_i \gamma_5 \tau_{\pi}(k) \right. \nonumber \\ 
&+& \left. i \left(\tau_{\pi}(p'-k) \Lambda_{\pi}^{\mu} \tau_{\pi}(p-k)\right)
\gamma_5 S(k) \gamma_5 \right]\,. \label{jq}
\end{eqnarray}

\begin{figure}[ht]
\begin{center}
\includegraphics[scale=1.0]{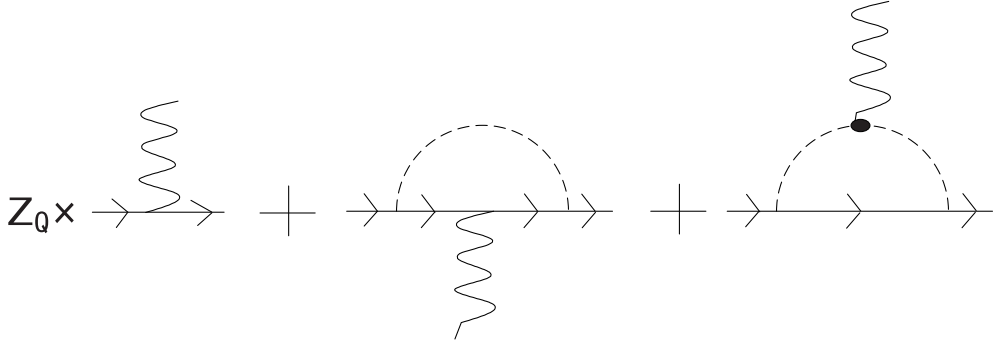}
\caption{Pion cloud contributions to the quark electromagnetic vertex.
$Z_Q$ is the quark wave function renormalization factor, and the dashed
line represents the t-matrix in the pion channel. (This figure actually
refers to the expression (\ref{jq1}), where the factor $Z_Q$ remains only
for the ``bare'' term.)}
\end{center}
\end{figure}

Here $\tau_{\pi}$ is the reduced $q{\overline q}$ t-matrix is
the pion channel, which can be approximated in the same way as
the diquark t-matrix (\ref{pole}):
\begin{eqnarray}
\tau_{\pi}(k) &=& \frac{-2i G_{\pi}}{1+2 G_{\pi}\Pi_{\pi}(k^2)}
\longrightarrow -2i G_{\pi} + \frac{i g_{\pi}}{k^2-M_{\pi}^2}
\equiv -2i G_{\pi} + i g_{\pi} \Delta_{F\pi}(k)\,. \nonumber \\
\label{piprop}
\end{eqnarray}
The bubble graph in the pion channel is $\Pi_{\pi}=\Pi_s$
(see (\ref{bubbs})), and the residue $g_{\pi}$ will be given in 
Eq.(\ref{gpi}) below. 

The expression (\ref{jq}) is formally similar to the nucleon current
(\ref{curr}). The presence of the first term simply expresses
the fact that in the NJL model the (bare) quarks are present from
the beginning, and the factor $Z_Q$ gives the probability of having a
constituent quark without its pion cloud. It has been expressed
in Eq.(\ref{self}) in terms of the self energy
\begin{eqnarray}
\Sigma_Q(p) = -3 \int \frac{{\rm d}^4 k}{(2\pi)^4} \left(\gamma_5
S(p-k) \gamma_5\right) \tau_{\pi}(k)\,.  \label{sigmaq}
\end{eqnarray}    

The quark electromagnetic vertex ${\Lambda}_{Q0}^{\mu}$ in (\ref{jq})
will be approximated by its point form after processing the renormalization
factor $Z_Q$, and the pion electromagnetic vertex $\Lambda_{\pi}^{\mu}
\equiv \tau_i \Lambda_{\pi, ij}^{\mu} \tau_j$ is similar to the diquark
vertex of Fig.2, but with point quark-photon couplings, as will be
specified below.

We now follow the same steps as for the calculation of the nucleon
current: We use the on-shell approximation for the quark and
pion vertices
\begin{eqnarray}
\left(S(\ell') \, \Lambda_{Q0}^{\mu}\, S(\ell)\right) &\longrightarrow&
Z_Q \left(S_F(\ell') \, \hat{\Lambda}_{Q0}^{\mu} \, S_F(\ell)\right) 
\label{pole0q} \\
\left(\tau_{\pi}(\ell') \, \Lambda^{\mu}_{\pi} \, \tau_{\pi}(\ell)\right) 
&\longrightarrow&
- g_{\pi} \left(\Delta_{F\pi}(\ell') \hat{\Lambda}_{\pi}^{\mu} \, 
\Delta_{F\pi}(\ell) \right) \,,
\label{polepi}
\end{eqnarray}
where the on-shell vertices are defined by 
\begin{eqnarray}
\hat{\Lambda}_{Q0}^{\mu} &\equiv& Z_Q \left(\Lambda_{Q0}^{\mu}\right)_
{\rm o.s.} \equiv
\gamma^{\mu} \, Q_Q \label{quark0} \\   
\hat{\Lambda}_{\pi,ij}^{\mu} &\equiv& g_{\pi} \left(\Lambda_{\pi,ij}^{\mu}
\right)_{\rm o.s.}
\equiv (\ell'+\ell)^{\mu} \left(-i \epsilon_{ij3}\right)
F_{\pi}(q^2)\,. \label{pionff}
\end{eqnarray}
Using $S(k)=Z_Q S_F(k)$ in the expression for $\Pi_{\pi}$, we get
\begin{eqnarray}
g_{\pi}^{-1} = \left( - \frac{\partial \Pi_{\pi}(q)}{\partial q^2}
\right)_{q^2=M_{\pi}^2} = Z_Q^2 \, 
\left( - \frac{\partial \hat{\Pi}_{\pi}(q)}{\partial q^2} \right)_
{q^2=M_{\pi}^2} \equiv Z_Q^2\,\hat{g}_{\pi}^{-1}\,,  
\label{gpi}
\end{eqnarray}
with the renormalized bubble graph 
$\hat{\Pi}_{\pi}=\hat{\Pi}_s$, see (\ref{newbubbs}).
Then Eq. (\ref{jq}) becomes
\begin{eqnarray}
\hat{\Lambda}_Q^{\mu} &=& Z_Q \gamma^{\mu} Q_Q
-i \hat{g}_{\pi} \frac{1}{2}(1-\tau_3) \int \frac{{\rm d}^4 k}{(2\pi)^4} 
\gamma_5 \left(S_F(p'-k) \gamma^{\mu} S_F(p-k) \right) 
\gamma_5 \Delta_{F\pi}(k) \nonumber \\ 
&-&  2i \, \tau_3\, \hat{g}_{\pi}\,  
F_{\pi}(q^2) \int \frac{{\rm d}^4 k}{(2\pi)^4}
\left(\Delta_{F\pi}(p'-k) \left(p'+p-2k\right)^{\mu} \Delta_{F\pi}(p-k)\right)
\gamma_5 S_F(k) \gamma_5\,.  \nonumber \\  \label{jq1}
\end{eqnarray}
Here we note that the contribution of the contact term ($2iG_{\pi}$)
to the quark diagram has been dropped in order to avoid double
counting: Because we always assume that our interaction Lagrangians
are Fierz symmetric, it is easy to see that this contribution
can be incorporated into the vector meson channel, which will be
separately considered below. By using $S = Z_Q S_F$ in the self energy
(\ref{sigmaq}) and in the expression for $Z_Q$ of Eq.(\ref{self}), 
we see that
\begin{eqnarray}
Z_Q =  1 + \left(\frac{\partial \hat{\Sigma}_Q}{\partial \fslash{p}}
\right)_{\fslash{p}=M}\,,  \label{zq2}
\end{eqnarray}
where the renormalized quark self energy is given by
\begin{eqnarray}
\hat{\Sigma}_Q(p) = -3 i \hat{g}_{\pi} \int 
\frac{{\rm d}^4 k}{(2\pi)^4} \left(\gamma_5
S_F(p-k) \gamma_5\right) \Delta_{F\pi}(k)\,.  \label{sigmaq1}
\end{eqnarray}    
The further evaluation of the loop integral (\ref{jq1}) is left to
Appendix B, where the contributions to the quark form factors
$F_{1Q}$ and $F_{2Q}$ are given. 
The pion electromagnetic vertex is evaluated from the definition 
(\ref{pionff}) and a Feynman diagram similar to Fig.2 for an external
pion, but with a point quark-photon coupling: 
\begin{eqnarray}
\hat{\Lambda}^{\mu}_{ij} = \left(-i \, \epsilon_{ij3}\right)\,
6i\, \hat{g}_{\pi}
\int \frac{{\rm d}^4k}{(2\pi)^4} {\rm Tr}_D \left[
\gamma_5 S_F(p'+k) \gamma^{\mu} S_F(p+k) \gamma_5 S_F(k) \right]\,. 
\label{lpi}
\end{eqnarray}
The explicit form of $F_{\pi}(q^2)$ is given in Appendix B.

Finally, we consider the corrections of the quark electromagnetic
vertex arising from vector mesons, similar to
vector meson dominance (VMD) models. If our original interaction Lagrangian
contains terms of the form
\begin{eqnarray}
{\cal L}_{I, v} = 
-G_{\omega} \left(\overline{\psi} \gamma^{\mu} \psi\right)^2
-G_{\rho} \left(\overline{\psi} \gamma^{\mu} {\bold \tau} \psi\right)^2\,,
\label{vectint}
\end{eqnarray}
then the point-like quark-photon vertices in the diagrams of Fig.3
and the pion vertex $\Lambda_{\pi}^{\mu}$ are replaced by the VMD 
vertex shown in Fig.4. 

\begin{figure}[ht]
\begin{center}
\includegraphics[scale=1.0]{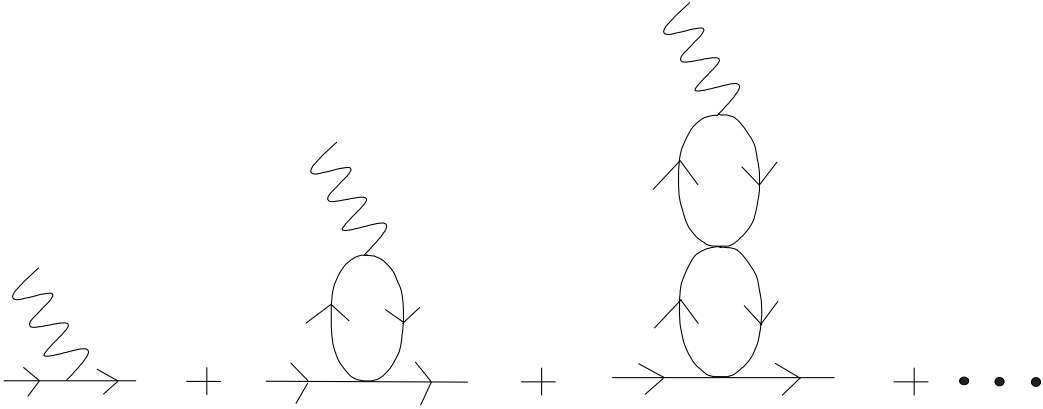}
\caption{The corrections to the quark electromagnetic vertex arising
from vector meson dominance. Here the 4-fermi interactions refer 
to the vector channel (see Eq.(\ref{vectint})), i.e., terms with tensor 
coupling to the external quark are not included.}
\end{center}
\end{figure}

Because of the transverse
structure of the bubble graphs in the vector channel, this 
leads to the following renormalization:
\begin{eqnarray}
\gamma^{\mu} Q_Q &\longrightarrow& \frac{1}{6} \left[
\gamma^{\mu} - \frac{2 G_{\omega} \hat{\Pi}_V(q^2)}{1+2 G_{\omega} 
\hat{\Pi}_V(q^2)}
\left(\gamma^{\mu} - \frac{q^{\mu}\fslash{q}}{q^2}\right) \right]
\label{is} \\
&+& \frac{\tau_3}{2} \left[
\gamma^{\mu} - \frac{2 G_{\rho} \hat{\Pi}_V(q^2)}
{1+2 G_{\rho} \hat{\Pi}_V(q^2)}
\left(\gamma^{\mu} - \frac{q^{\mu}\fslash{q}}{q^2}\right) \right]\,,
\label{iv} 
\end{eqnarray}
where the form of $\hat{\Pi}_V$ is given in Appendix B.
Because our quark-photon vertex in (\ref{quarkff}) is defined for on-shell
quarks, the terms ${\displaystyle \propto \frac{q^{\mu}\fslash{q}}{q^2}}$ 
do not
contribute. Therefore the isoscalar
(or isovector) parts in the quark electromagnetic vertex 
should be multiplied by a form factor
\footnote{Because the scalar diquark has isospin zero, this 
eventually also holds for the nucleon electromagnetic vertex.}
$F_{\omega}(q^2)$ (or $F_{\rho}(q^2)$), where
\begin{eqnarray}
F_{\alpha}(q^2) = \frac{1}{1+2 G_{\alpha} \hat{\Pi}_V(q^2)}
\,\,\,\,\,\,\,(\alpha=\omega,\rho)\,.
\label{vmd}
\end{eqnarray}

\section{Results for the nucleon form factors}
\setcounter{equation}{0}

In this section we will show our results for the nucleon
form factors. First we discuss our model parameters and particle
masses. The parameters are the same as in Refs.\cite{BT,QM}:
The IR cut-off $\Lambda_{\rm IR}$ is fixed as $0.2$ GeV, 
the constituent quark mass at zero nucleon density is $M_0=0.4$ GeV,
and $G_{\pi}$, $\Lambda_{\rm UV}$ are
determined so as to reproduce $M_{\pi}=0.14$ GeV 
and $f_{\pi}=93$ MeV. This gives $G_{\pi}=19.60$ GeV$^{-2}$
and $\Lambda_{\rm UV}=0.6385$ GeV.
The coupling constant $G_s$ is
determined so as to reproduce $M_{N0}=0.94$ GeV, 
which gives the ratio $G_s/G_{\pi}=0.508$.
The coupling constant $G_{\omega}$ is
determined so that the curve for the NM
binding energy per nucleon $(E_B/A$) as a function of the density
passes through the empirical saturation
point\footnote{We recall from Ref.\cite{BT} that, in this simple NJL model, we
cannot adjust both the empirical binding energy and saturation density
at the same time. 
Therefore, although the binding energy curve passes through the 
empirical saturation point, its minimum is at a different point,
($\rho, E_B/A)=(0.22$ fm$^{-3},\,17.3$ MeV).} 
($\rho, E_B/A)=(0.16$ fm$^{-3},\,15$ MeV),
which gives the ratio $G_{\omega}/G_{\pi}=0.37$.
Finally, for the VMD form factors (\ref{vmd}) we also need the
coupling constant $G_{\rho}$, which is determined by reproducing the
empirical symmetry energy coefficient $a_4=35$ MeV at the density 
$\rho=0.16$ fm$^{-3}$, which gives $G_{\rho}/G_{\pi}=0.091$

In Table 1 we list the effective quark,
diquark, nucleon and pion masses for the densities $\rho=0,\,0.08,\, 
0.16,$ and $0.24$ fm$^{-3}$. Concerning the
pion mass in the medium, we use a general result based on chiral
symmetry \cite{SIGMA}, which for the NJL model implies that the 
product $M_{\pi}^2\,\cdot M$ is a constant independent of density,
see Eq.(2.58) of Ref.\cite{BT}. We therefore use
\footnote{More precisely, this pion mass is defined at zero momentum,
and includes nucleonic (Z-graph and contact) terms, which are
important in order to guarantee the Goldstone nature of the pion in
the medium. We note that these nucleonic contributions to the scalar
diquark (or sigma) mass at normal densities are numerically small
compared to the $qq$ (or $q{\overline q}$) polarizations, 
although they become important for small $M$ and guarantee the 
stability of the system w.r.t. variations in $M$, see Ref.(\cite{BT}) 
for details.} $M_{\pi}^2 = M_{\pi 0}^2 \cdot M_0/M$. Also listed
in Table 1 are the values of $\hat{g}_s$, $\hat{g}_{\pi}$, $\hat{Z}_N$
and $Z_Q$.     

\begin{table}[h]
\begin{center}
\begin{tabular}{|c|c|c|c|c|}
\hline
&  $\rho=0$ & $\rho=0.08\,$fm$^{-3}$ & $\rho=0.16\,$fm$^{-3}$ & 
$\rho=0.24\,$fm$^{-3}$  \\  \hline
$M$               &  0.4   &  0.353  &  0.308  & 0.268  \\
$M_s$             &  0.576 &  0.493  &  0.413  & 0.342  \\
$M_N$             &  0.94  &  0.818  &  0.707  & 0.619  \\
$M_{\pi}$         &  0.14  &  0.149  &  0.159  & 0.171  \\
$\hat{g}_s$       &  18.26 &  17.17  &  16.20  & 15.40  \\
$\hat{g}_{\pi}$   &  17.81 &  14.29  &  11.60  & 9.70   \\ 
$\hat{Z}_N$       &  44.32 &  47.62  &  50.38  & 51.97  \\ 
$Z_Q$             &  0.808 &  0.832  &  0.852  & 0.869  \\  \hline 
\end{tabular}
\end{center}
\caption{Effective masses for the quark, the diquark, the nucleon and the
pion (all in GeV), and pole residues $\hat{g}_s$, $\hat{g}_{\pi}$, 
$\hat{Z}_N$ and $Z_Q$ for four values of the density.}  
\end{table}

We will discuss our results for the nucleon form factor in terms of 
the Dirac-Pauli form factors defined by Eq.(\ref{par1}). In the 
discussion of medium effects, in particular the effects of the
reduced nucleon mass ($M_N < M_{N0}$), we will also refer to an
equivalent parametrization in terms of the ``orbital form factor'' 
$G_{L}$ and the ``spin form factor'' $G_S$:
\begin{eqnarray}
j_N^{\mu} &=& \sqrt{\frac{M_N}{E_{Np}}\,\frac{M_N}{E_{Np'}}} 
\, \overline{u}_N(p') \left[ \frac{\left(p'+p\right)^{\mu}}
{2 M_{N0}} G_{L}(q^2)
+ \frac{i \sigma^{\mu \nu}q_{\nu}}{2M_{N0}} G_S (q^2) \right] u_N(p) 
\nonumber \\
\label{par2}
\end{eqnarray}
The relations to the Dirac-Pauli form factors are
\begin{eqnarray}
G_{L} = \frac{M_{N0}}{M_N} F_{1N}\,,\,\,\,\,\,\,\,\,\,\,
G_S = \frac{M_{N0}}{M_N} (F_{1N} + F_{2N})\,. \label{rel}
\end{eqnarray}
We introduce these form factors here, because $F_1$ and
$F_2$ do not directly reflect the enhancement of the nucleon orbital
current (${\bold j}_{\rm N, orb}$) arising from the reduced nucleon 
mass (enhanced nuclear magneton)
\footnote{Note that the appearance of the medium modified nucleon mass
($M_N$) in the Pauli term of (\ref{par1}) is a mere definition of 
$F_2$.}. Moreover, the parametrization (\ref{par2}) for the space
part of the current 
has more connection to the traditional calculations of nuclear magnetic 
properties \cite{ASBH}, because the values of these form factors at 
$q=0$ reduce to the
orbital and spin g-factors: $g_{L}=G_L(0)$, $g_{S}=G_S(0)$.  

Here we would like to point out that these different ways to discuss 
the medium modifications of the nucleon
current remind us that the form factors of a nucleon in the medium are not
directly observable quantities: Ultimately the current $j_N^{\mu}$
has to be used in a nuclear structure calculation of observable cross 
sections. 
Our current $j_N^{\mu}$ reflects only those effects which are not taken
into account in nuclear structure calculations, i.e., the effects of the
nuclear mean fields on the internal motion of quarks in the nucleon. 
Other effects, which explicitly depend on the density and have their
origin in the Pauli principle on the level of nucleons, must be 
considered in the nuclear part of the calculation
\footnote{This is also evident from
the fact that the full current of a nucleon in the medium, including the
explicitly density dependent parts, cannot be parametrized in
the Lorentz invariant forms (\ref{par1}) or (\ref{par2}).}. 
As an example of such a calculation for the case of nuclear matter, we
will consider the response function for quasielastic electron scattering
in Sect.5.
 
For the zero density (single nucleon) case, it is possible to define 
a Breit frame where $q_0=0$, and in this frame the nucleon current can be 
expressed in terms of the familiar electric and magnetic form factors
\begin{eqnarray}
G_E(q^2) &=& F_1(q^2) + \frac{q^2}{4 M_N^2} F_2(q^2) \label{ge} \\
G_M(q^2) &=& F_1(q^2) + F_2(q^2)\,. \label{gm}
\end{eqnarray}
We will compare our calculated form factors for zero density with the 
empirical dipole form factors, defined by
\begin{eqnarray}
G_{Ep} &=& G_{Mp}/\mu_p = G_{Mn}/\mu_n = 
\frac{1}{\left(1-q^2/ 0.71\,{\rm GeV}^2\right)^2} \nonumber \\     
F_{1n} &=& 0\,.  \label{dipole}
\end{eqnarray}
For a nucleon moving in the medium, however, one cannot define
a Breit frame, and consequently the combinations (\ref{ge}) and (\ref{gm}) 
do not enter naturally in the expressions for nuclear observables, 
like response functions or elastic form factors. For the
finite density case we will therefore discuss our results in terms
of the form factors $F_1$ and $F_2$, or $G_L$ and $G_S$.  

The results for the form factors at zero density (free nucleon case) are
shown in Figs.5-8. There we plot (i) the results of the naive quark-diquark
model (see Sect.3.1) without (dotted lines) and with (dashed lines) the 
contact term contribution to the quark diagram of Fig.1, (ii) the results
obtained by including in addition the effects of the diquark form
factor (dash-dotted lines), and (iii) the total result including also 
the pion and VMD effects.

\begin{figure}[ht]
\begin{center}
\includegraphics[scale=0.7]{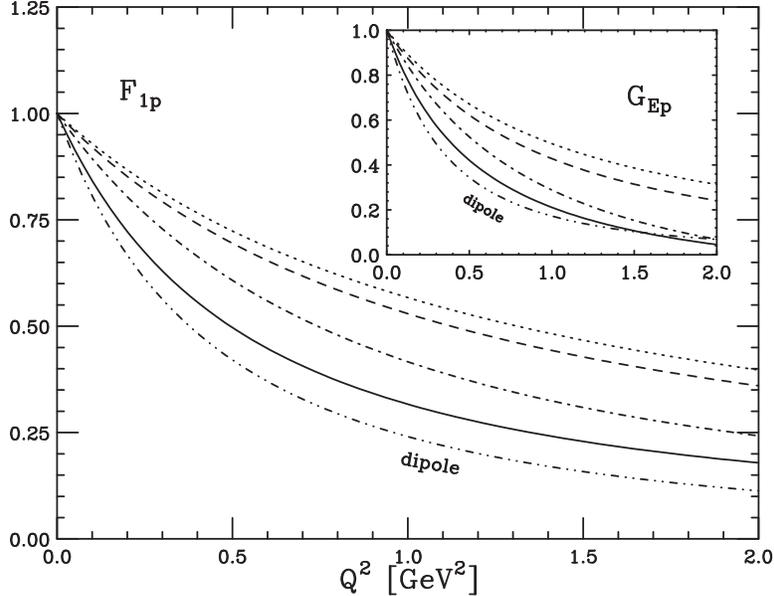}
\caption{The form factor $F_{1p}$ for zero density. The dotted line
is the result of the naive quark-diquark model without the contact term, 
to which the other contributions are successively added as follows:
Dashed line: including contact term; dash-dotted line: including diquark
form factor; solid line: full results including pion and vector meson
contributions. The dash-double dotted line labeled ``dipole'' shows the
dipole parametrization. The insert shows the form factor $G_{Ep}$, 
with the same meaning of the lines.}
\end{center}
\end{figure}

Fig. 5 shows that in the naive quark-diquark model the electric size
of the proton is too small and the form factors $F_{1p}$ and $G_{Ep}$ 
fall off too slowly. The situation improves
when the diquark form factor is included, and also the pion cloud
gives some positive contribution to the electric size of the proton
\footnote{We obtain $<r^2_E>_{p}=0.421\,$fm$^2 + 0.062\,$fm$^2 = 0.483$ 
fm$^2$, where the two terms come from the Dirac ($F_1$) part and the anomalous 
($F_2$) part. The fact that this is too small compared to the 
experimental value of $0.74\, $fm$^2$ is partially because the magnetic moment
is too small, but also because the slope of $F_1$ is too small.}.  
The total result for $F_{1p}$ still lies above the empirical dipole
form factor, but we can expect that the further inclusion of axial
vector diquarks and pion cloud effects around the {\em nucleon} will
lead to a satisfactory description. 

\begin{figure}[ht]
\begin{center}
\includegraphics[scale=0.7]{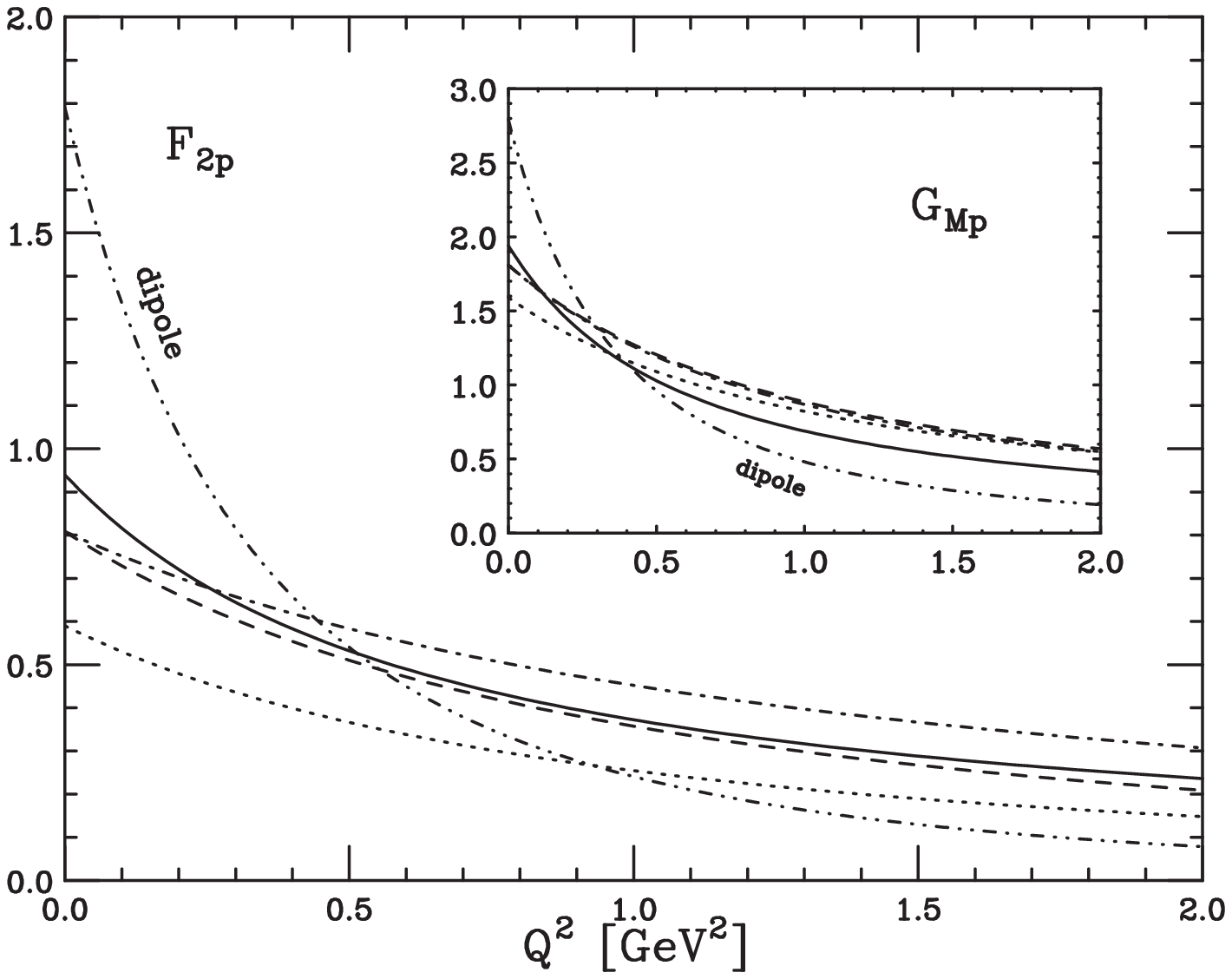}
\caption{Same as Fig. 5 for the form factors $F_{2p}$ (main part)
and $G_{Mp}$ (insert) for zero density.}
\end{center}
\end{figure}

Fig. 6 shows that the proton magnetic
moment in the naive quark-diquark picture is too small, which
is expected and also well known\cite{AX,OETT}. The finite size effects 
of the scalar
diquark do not contribute much in this case. The inclusion of the contact term
in the quark diagram of Fig.1 improves the situation somewhat. By using
a Fierz transformation, this term is actually seen to be equivalent to
a vector meson contribution (like shown in Fig.4 for the quark), but with 
a tensor coupling ($\propto \sigma^{\mu \nu}q_{\nu}$) to the nucleon.
Also the pion cloud, which leads to anomalous magnetic moments of the
constituent up and down quarks\footnote{We obtain $\mu_u=\frac{2}{3}+0.061$,
$\mu_d=-\frac{1}{3}-0.123$ for the magnetic moments of u,d quarks in the
free nucleon.}, gives a positive contribution to
the proton magnetic moment, but the total result is still too small.
It is, however, known that the axial vector diquark and the pion
cloud around the {\em nucleon} give large contributions to the magnetic 
moment and the associated form factors,
and Fig.6 shows how far one can go in the simple quark - scalar 
diquark description. 

The importance of the diquark form factor is also seen for the neutron form
factors $F_{1n}$ and $G_{En}$, which are shown in Fig. 7. The naive
quark-diquark model gives an electric form factor which is too
large in comparison to the experimental one (note that the experimental 
$G_{En}$ is smaller than the ``dipole form factor'' shown in Fig. 7), 
and the diquark form factor,
which suppresses the (positive) contribution from the second diagram of
Fig.1 relative to the (negative) first one, is essential
to obtain reasonable values. The electric size of the neutron is
somewhat too small in magnitude, but this is because the absolute value
of the magnetic moment is too small
\footnote{We obtain $<r^2_E>_{n}=0.003\,$fm$^2 - 0.072\,$fm$^2 = - 0.069$ 
fm$^2$, where the two terms come from the Dirac ($F_1$) part and the anomalous 
($F_2$) part. The fact that this is too small in magnitude compared to the 
experimental value of $-0.12\, $fm$^2$ is because the absolute value of the
magnetic moment
is too small. For discussions on the role of these two contributions to
the neutron electric radius, see for example \cite{ISG}.}. 
In this connection, it is interesting to observe that 
the result for $F_{1n}$, and in particular its contribution to the 
electric radius, is very small, and therefore the electric size of the
neutron is almost entirely due to the ``Foldy term'' \cite{ISG}.     
 
\begin{figure}[ht]
\begin{center}
\includegraphics[scale=0.7]{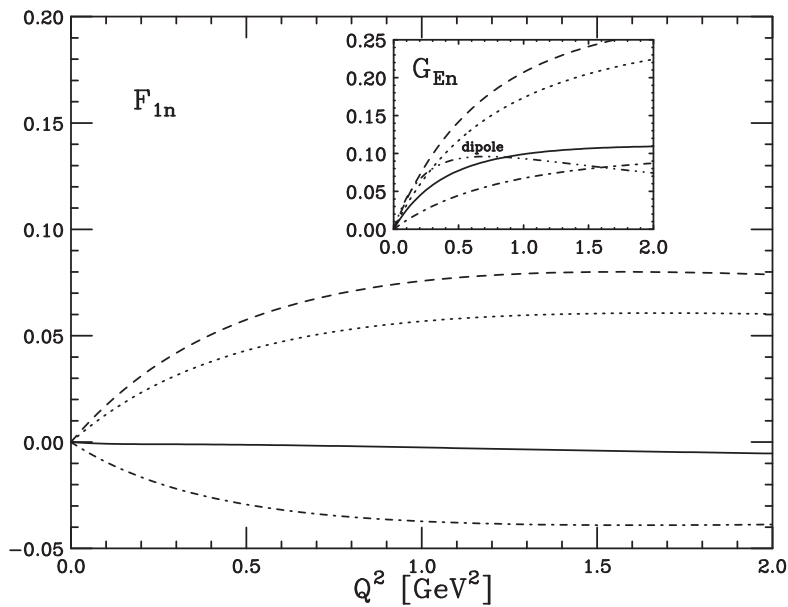}
\caption{Same as Fig. 5 for the form factors $F_{1n}$ (main part)
and $G_{En}$ (insert) for zero density. The dipole parametrization of
$F_{1n}$ is zero by definition and therefore not indicated here.}
\end{center}
\end{figure}

The situation for the form factors
$F_{2n}$ and $G_{Mn}$ shown in Fig. 8 is similar to the case of 
the proton (Fig. 6), i.e., the contact term and the pion cloud around
the constituent quarks give some improvements of the magnetic moment,
but the total result is still too small. This, and the fact that
the form factors fall off too slowly, again points out the
necessity to include the axial vector diquark channel.

\begin{figure}[ht]
\begin{center}
\includegraphics[scale=0.7]{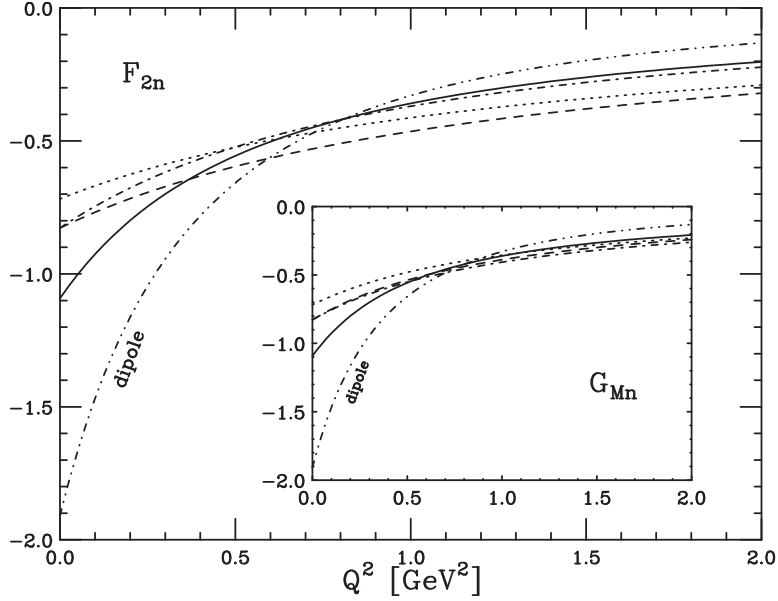}
\caption{Same as Fig. 5 for the form factors $F_{2n}$ (main part)
and $G_{Mn}$ (insert) for zero density.}
\end{center}
\end{figure}

The medium modifications of the nucleon form factors are shown in
Figs. 9-12, where we plot the results for $\rho=0$ (dotted lines),
$\rho=0.08$ fm$^{-3}$ (dashed lines), $\rho=0.16$ fm$^{-3}$ (solid lines),
and $\rho=0.24$ fm$^{-3}$ (dash-dotted lines). 

\begin{figure}[ht]
\begin{center}
\includegraphics[scale=0.7]{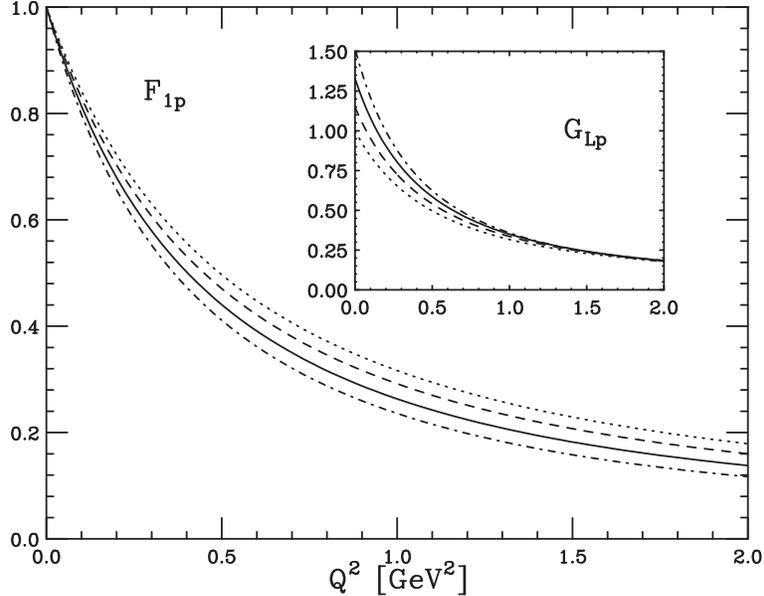}
\caption{The form factor $F_{1p}$ for the cases $\rho=0$ (dotted line),
$\rho=0.08$ fm$^{-3}$ (dashed line), $\rho=0.16$ fm$^{-3}$ (solid line), 
$\rho=0.24$ fm$^{-3}$ (dash-dotted line). The insert shows the proton
orbital form factor $G_{Lp}$ defined in Eq.(\ref{par2}) with the same meaning
of the lines.}
\end{center}
\end{figure}

The result for $F_{1p}$ of
Fig. 9 indicates that the electric size of the proton in the medium
is somewhat enhanced. 
The orbital form factor $G_{Lp}$ shown in the insert of Fig.9 
demonstrates the enhancement of the orbital current 
(${\bold j}_{\rm N, orb}$) due to the
reduced effective nucleon mass, see Eq.(\ref{rel}). We have to
remind, however, that the isoscalar part of this enhancement is
in a sense spurious, because in an actual nuclear calculation it is 
canceled by the ``backflow'' effect\cite{BACK}, which in our language arises 
from Z-graphs, i.e., the Pauli blocking part of the $N\overline{N}$ 
excitation piece (see the detailed discussions in Ref.\cite{ASBH,BACK1} 
on the backflow in relativistic meson-nucleon theories). 
Namely, the proton orbital g-factor, which is roughly $M_{N0}/M_N$ in a Hartree
calculation, becomes approximately $\frac{1}{2} (1+M_{N0}/M_N)$ 
after the inclusion of the backflow, where the first term is the
isoscalar and the second one the isovector piece\footnote{It is well
known that the pion effects further enhance the isovector piece\cite{ASBH}.}. 
 
\begin{figure}[ht]
\begin{center}
\includegraphics[scale=0.7]{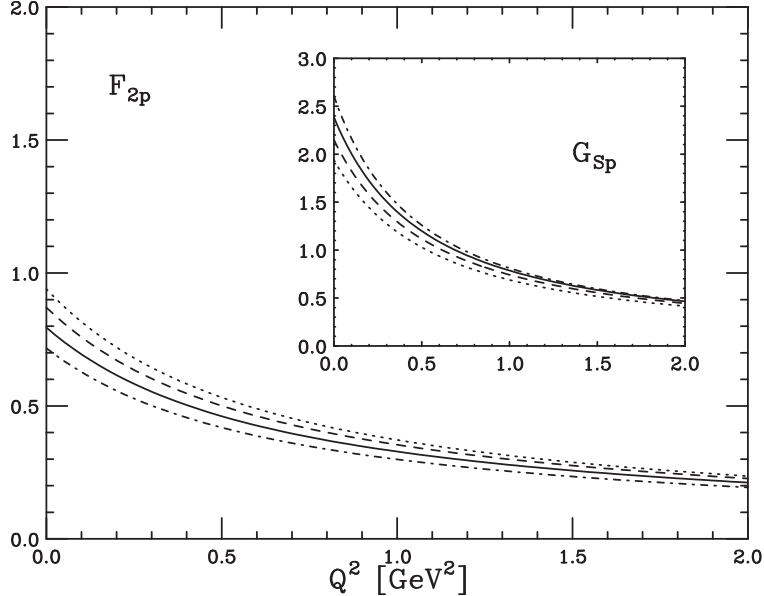}
\caption{Same as Fig.9 for the form factor $F_{2p}$ (main part)
and the proton spin form factor $G_{Sp}$ defined in Eq.(\ref{par2}) 
(insert).}
\end{center}
\end{figure}

Fig. 10 shows that the medium effects tend to decrease the ``intrinsic''
anomalous magnetic moment of the proton, but when combined
with the enhancement of the nuclear magneton, the spin g-factor is
enhanced, as shown in the insert of Fig.10. It is interesting
that a very similar result has been obtained also in
hadronic models\cite{BAB}. Therefore, the quark effects considered here
do not lead to a quenching of the spin g-factor, as would be desirable
to explain the missing quenching of isovector nuclear magnetic moments
\cite{ASBH}, 
but rather to an enhancement
\footnote{This is in contrast to the quenching of the 
axial vector coupling constant observed in the quark-diquark calculations of
Ref.\cite{NEWI} including the axial vector diquark, and in hadronic models
\cite{BAB1}.}. The figure also
shows that the magnetic size of the proton becomes somewhat larger
in the medium.   

The results for the neutron form factor $F_{1n}$ in Fig. 11
show that the effect of finite diquark size, which was very important
for the zero density case (Fig.7) to reduce $F_{1n}$ to
reasonable values, increases with increasing density. That is,
the diquark form factor at finite density
further suppresses the positive contribution of the diquark diagram
in Fig.1. The orbital form factor $G_{Ln}$ shown in the insert of
Fig.11 again demonstrates the enhancement of the nuclear magneton,
but we have to keep in mind that the backflow effect will change
the neutron orbital g-factor from the present value $0$ to roughly
$\frac{1}{2}\left(1-M_{N0}/M_N\right)<0$, and that effects of the
pion cloud around the nucleon are known to further enhance 
the magnitude of the neutron orbital g-factor.

\begin{figure}[ht]
\begin{center}
\includegraphics[scale=0.7]{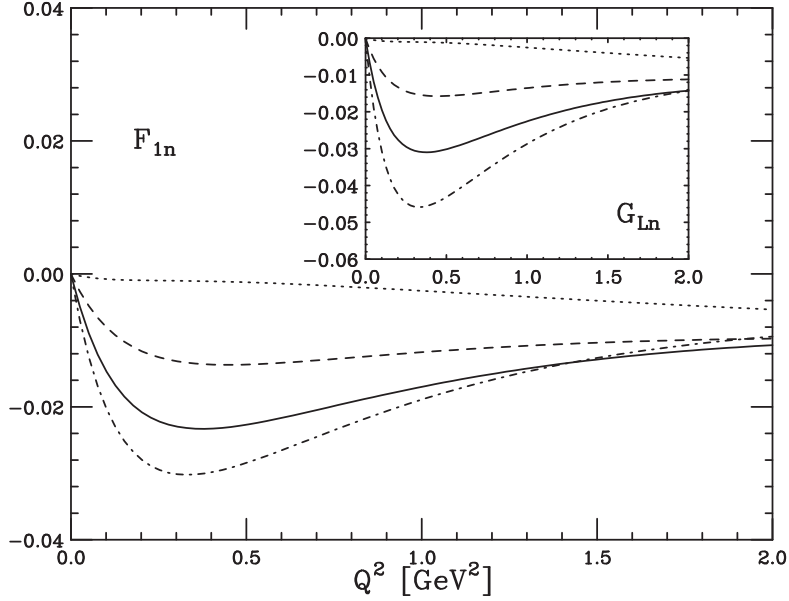}
\caption{Same as Fig.9 for the form factor $F_{1n}$ (main part)
and the neutron orbital form factor $G_{Ln}$ defined in 
Eq.(\ref{par2}) (insert).}
\end{center}
\end{figure}

The medium effects on the neutron form factor $F_{2n}$ shown in
Fig.12 are qualitatively similar to the proton case of Fig.10:
They decrease the absolute value of the ``intrinsic''
anomalous magnetic moment of the neutron, but when combined
with the enhancement of the nuclear magneton the neutron 
spin g-factor becomes slightly enhanced in magnitude, as shown 
in the insert of Fig.12.
The magnetic size of the neutron is not changed much, and the total 
medium effects on the neutron spin form factor are very small.

An interesting common feature of the results shown in Fig. 9-12 is that
the medium modifications of the orbital and spin form factors
($G_L(Q^2)$ and $G_S(Q^2)$) always decrease with increasing $Q^2$. This is
consistent with the intuitive expectation that the internal structure
of the nucleon at short distances is not influenced much by the
mean nuclear fields, and again indicates that these form factors
reflect the change of the nucleon current in the medium more directly
and transparently than the form factors $F_{1}$, $F_2$ themselves, 
or any other combination of them.  

\begin{figure}[ht]
\begin{center}
\includegraphics[scale=0.7]{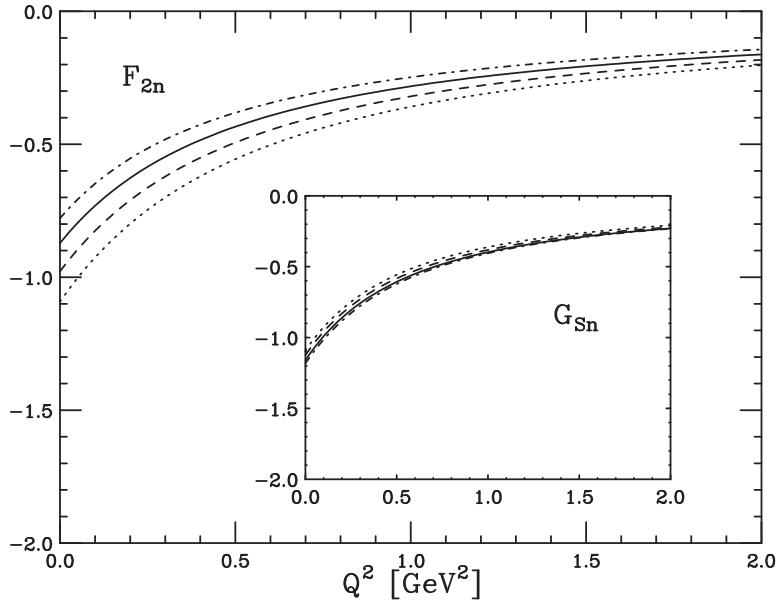}
\caption{Same as Fig.9 for the form factor $F_{2n}$ (main part)
and the neutron spin form factor $G_{Sn}$ defined in 
Eq.(\ref{par2}) (insert).}
\end{center}
\end{figure}

\section{Application: The longitudinal nuclear response function}
\setcounter{equation}{0}

As an application of the medium effects discussed in the previous
section to the calculation of nuclear quantities, we consider the
longitudinal response function for quasielastic electron scattering.
Generally, quasielastic processes are the ideal places to investigate
medium modifications of nucleon form factors, because the 
on-shell kinematics, which is used in the derivation of the nucleon
form factors, is justified. Recently, experiments on the
polarization transfer in proton knock-out reactions have been carried
out in the region of quasielastic kinematics at large momentum
transfers\cite{POL}, and the results were discussed in connection to
the predicted medium
modifications of nucleon form factors\cite{MEDF}. In this work, however, we
will consider the simpler case of the inclusive quasielastic
response function in the region of lower momentum transfers. In
previous works\cite{WAL}, based on purely hadronic models, it was shown that
this quantity is quite sensitive to medium modifications of form
factors, and here we wish to apply our more microscopic quark description
to the longitudinal response function in nuclear matter. 
  
The longitudinal response function in isospin symmetric nuclear matter
is expressed as
\begin{eqnarray}
S_L(\omega,{\bold q}) = \frac{2 Z}{\pi \rho} \, {\rm Im} \, 
\Pi_L(\omega,{\bold q})\,,
\label{resp}
\end{eqnarray}
where $\Pi_L(\omega,{\bold q})$ is the 2-point (correlation) 
function for two external operators $j_N^0$. In the mean field (Hartree)
approximation, this is expressed by the first diagram of Fig.13, and
if we include the direct terms of the NN interaction in the ladder
approximation, this gives the familiar RPA series. 

\begin{figure}[ht]
\begin{center}
\includegraphics[scale=1.0]{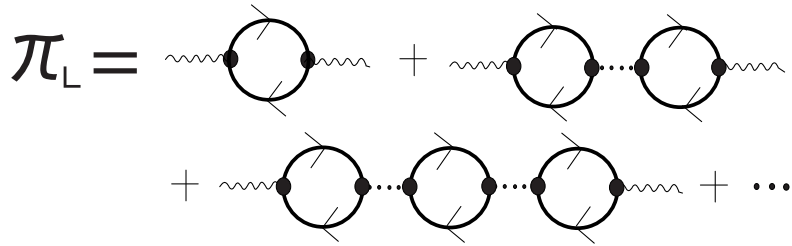}
\caption{Graphical representation of the longitudinal current-current
correlation function. The solid lines represent nucleons, and the
dashed lines the interaction between nucleons. The external operator
is $j_N^0$. The first diagram gives the Hartree response function, and
the others the RPA corrections.}
\end{center}
\end{figure}

The relativistic calculation of $\Pi_L$ follows the lines given in 
Ref.\cite{KS,WEHR}, i.e., the
bubble graphs in Fig.13 consist of the nucleon particle- nucleon hole 
excitations and the Pauli blocking part of the nucleon particle- antinucleon
excitations (Z-graphs). The other effects, which are not taken into account
explicitly in Fig.13, are summarized in the density dependent electromagnetic
vertices. Previous calculations based on hadronic models\cite{WAL} incorporated
the vacuum fluctuations on the level of nucleons (that is, the change of
the nucleon-antinucleon vacuum polarization
graphs in the presence of the nuclear mean fields), 
but it is more appropriate to describe
these vacuum fluctuations on the level of quarks
\footnote{Although this follows naturally from the derivation of the nucleon
lagrangian from the quark lagrangian in the path integral approach
\cite{QM}, or from
the derivation of the effective NN interaction in quark theories following
the Landau-Migdal approach\cite{BT}, it remains to be demonstrated 
explicitly for electromagnetic quantities.}. Therefore we use our 
nucleon current (\ref{curr}) for $\mu=0$ at the electromagnetic 
vertices of Fig.13. 

For the RPA calculation, we need the NN interaction kernel in our effective
quark theory, which is shown graphically in Fig.14 and expressed as
\footnote{The quantities $({\bold 1})_i$ and $(\gamma^{\mu})_i$ 
in (\ref{sigma})
and (\ref{omega}) express the Dirac matrices acting between the spinors
of nucleon i=1,2, i.e.,
in order to get the NN interaction one has to multiply the spinors
$\sqrt{M_N/E_{N}} u_N$ of the initial and final states.}
\begin{eqnarray}
\lefteqn{V_{NN}(k) \equiv
V_{\sigma}(k) \, ({\bold 1})_1 \cdot ({\bold 1})_2
+ V_{\omega}(k) \, (\gamma_{\mu})_1 \cdot ({\gamma^{\mu}})_2}  \label{vnn} \\
& & = \frac{-2 G_{\pi}}{1 - 2 G_{\pi}\, \hat{\Pi}_{\sigma}(k^2)
+ 2 G_{\pi} \, \delta M_{s}^2} 
\left(\frac{{\rm d}M_N}{{\rm d}M} \right)^2 F_{\sigma}^2(k^2)\, 
({\bold 1})_1 \cdot ({\bold 1})_2 \label{sigma} \\
&+& \frac{2 G_{\omega}}{1 + 2 G_{\omega} \, \hat{\Pi}_V(k^2)} \, 9 \, 
F_{\omega}^2(k^2) \, (\gamma_{\mu})_1 \cdot ({\gamma^{\mu}})_2 
\label{omega} 
\end{eqnarray}

\begin{figure}[ht]
\begin{center}
\includegraphics[scale=1.0]{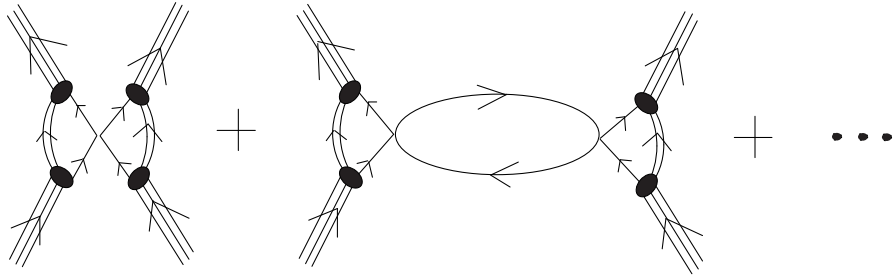}
\caption{The NN interaction of Eq.(\ref{vnn}) 
in the present quark-diquark 
model. Only those terms arising from the interaction of the two spectator
quarks are shown, there are also equivalent graphs from the interaction
between the quarks inside the diquarks. The term $\delta M_{\sigma}^2$,
which is included in Eq.(\ref{sigma}), is not shown explicitly in this figure.}
\end{center}
\end{figure}

Here the first factor in (\ref{sigma}) is the reduced $q{\overline q}$ 
t-matrix in the sigma meson channel, and the corresponding bubble graph
$\hat{\Pi}_{\sigma}$ 
is given in Appendix C. The derivation of the effective NN interaction
in the NJL model\cite{BT}, however, has shown that in addition to the part
$\hat{\Pi}_{\sigma}$, which describes the $q\overline{q}$ exchange
(see Fig.14), there is also a nuclear part which consists of (i)
the Z-graph, and (ii) a contact term arising from an induced
$\overline{N}\sigma^2 N$ interaction. Concerning the Z-graph, we
note that this is just the Pauli blocking part to the $N \overline{N}$
bubble graph, which is taken into account explicitly in the RPA series
of Fig.13 and therefore should not be included in the interaction.
(Numerically the Z-graph contribution is small compared to 
$\hat{\Pi}_{\sigma}$ because of the
reduced $\sigma$N coupling in the medium, see Ref.\cite{BT}.) 
The self energy correction for the sigma meson arising from the induced
$\overline{N}\sigma^2 N$ contact interaction is included in (\ref{sigma})
as the density-dependent constant\cite{BT} 
\begin{eqnarray}
\delta M_{\sigma}^2 = 4 \left(\frac{{\rm d}^2 M_N}{{\rm d}M^2}\right)
\int \frac{{\rm d}^3 k}{(2\pi)^3} \frac{M_N}{E_N(k)}\Theta(p_F-|{\bold k}|)\,.
\label{delta}
\end{eqnarray}
The $NN \sigma$ coupling constant at zero momentum is proportional to
${\rm d}M_N/{\rm d}M$, and its square appears in (\ref{sigma}). The
$NN\sigma$ vertex form factor $F_{\sigma}(k^2)$ is normalized to 
$F_{\sigma}(0)=1$. The first factor in (\ref{omega}) is the reduced 
$q{\overline q}$ 
t-matrix in the $\omega$-meson channel, and the corresponding bubble graph
$\hat{\Pi}_V$ is the same as in the VMD correction to the quark form factors
(Fig.4).
The $NN \omega$ coupling constant at zero momentum is proportional to
the number of quarks in the nucleon, and the
vertex form factor $F_{\omega}(k^2)$ is normalized as
$F_{\omega}(0)=1$.

\begin{figure}[ht]
\begin{center}
\includegraphics[scale=0.7]{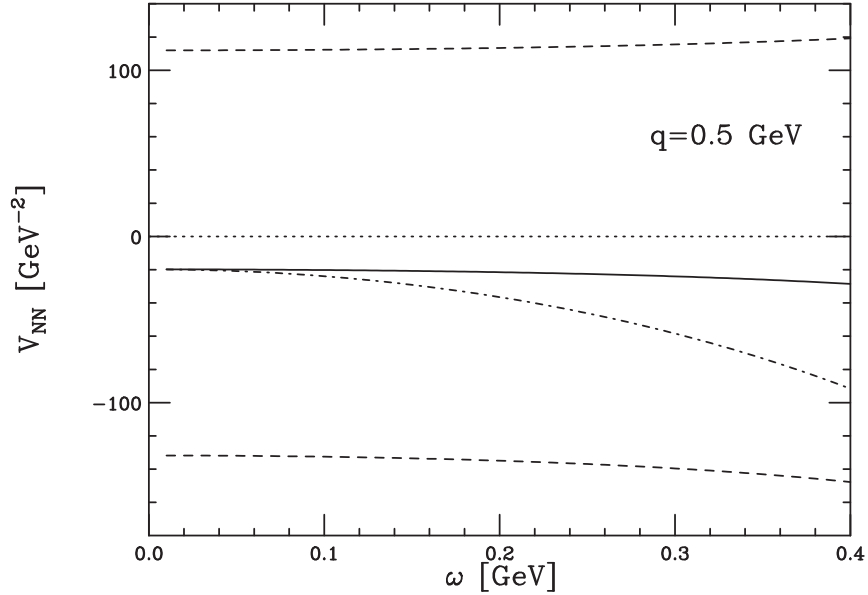}
\caption{The NN interaction for $|{\bold q}| = 0.5$
GeV as a function of $\omega$. The lower dashed line shows the attractive
part $V_{\sigma}$ of Eq.(\ref{sigma}), the upper dashed line the 
repulsive piece
$V_{\omega}$ of Eq.(\ref{omega}), the solid line shows their naive sum, 
and the dash-dotted line shows the combination 
$V_{\sigma}+V_{\omega}(1-\omega^2/|{\bold q}|^2)$.}
\end{center}
\end{figure}

By taking matrix elements of (\ref{sigma}), (\ref{omega}) between the
nucleon spinors, it is easy to see that for $k=0$ and for nucleons at the
Fermi surface we just get the
Landau-Migdal interaction derived more generally in Ref.(\cite{BT}) 
(except for the Z-graph contributions as explained above). 
The form factor $F_{\omega}$ is equal to
$(F_{1p}+F_{1n})$, which was calculated in
the previous sections. The scalar form factor $F_{\sigma}$ should in 
principle be calculated independently from the Feynman diagrams of Fig.1 
by using the external
operator ${\bold 1}$. However, because the calculations discussed below
show that the RPA effects are numerically not very important, we will
simply assume the same form factor as for the $NN \omega$ coupling
($F_{\sigma}=F_{\omega}$).

The two parts of the NN interaction, $V_{\sigma}$ of (\ref{sigma}) and 
$V_{\omega}$ of (\ref{omega}) ,
are shown by the lower and upper dashed lines, respectively, in Fig.15 
for the kinematics needed in the calculation of the response
function, $|{\bold q}|=0.5$ GeV and $0<\omega<0.4$ GeV.
As in the case of relativistic hadronic models\cite{SW}, we see large 
cancellations between the attractive scalar and repulsive vector parts. 
The solid line shows the naive
sum of the two dashed lines, while the dash-dotted line shows the
combination $V_{\sigma}+V_{\omega}(1-\omega^2/|{\bold q}|^2)$, which
includes the effect of the longitudinal space component of the $\omega$
exchange and is more relevant for the longitudinal response function
\cite{WAL}\footnote{We have to note, however, that the true interaction
in the nuclear medium in the longitudinal channel is more repulsive than 
shown by the dash-dotted line in Fig.15, because of the difference
between the Dirac matrices ${\bold 1}$ and $\gamma^0$ in Eq.(\ref{vnn}).
On the average it is repulsive on the small $\omega$ side and becomes 
attractive on the large $\omega$ side, as the RPA result of Fig.16 shows.}.  
The results shown in Fig. 15 can be reproduced almost exactly by
an approximate form in terms of Yukawa potentials, if the
coupling constants and meson masses are defined at $k^2=0$. This is
discussed in Appendix C, where also numerical values are given.

\begin{figure}[ht]
\begin{center}
\includegraphics[scale=0.7]{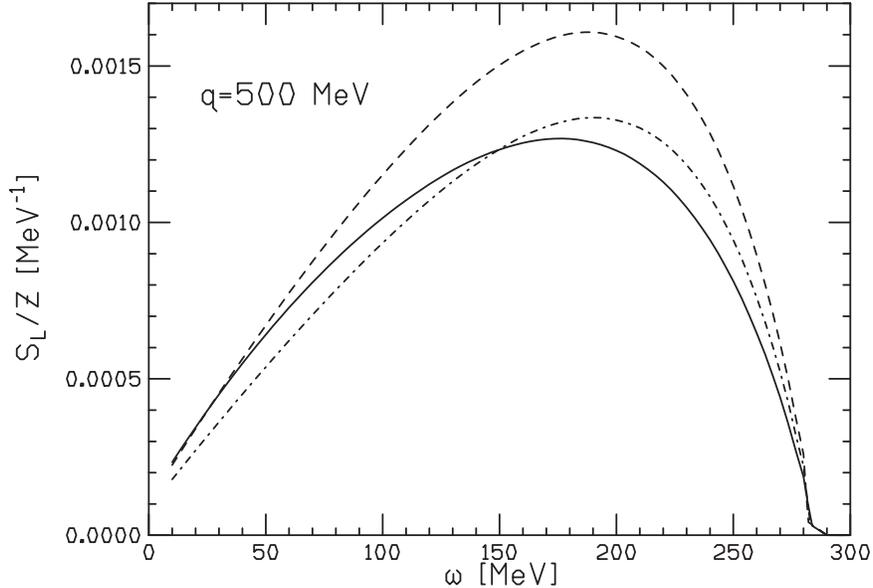}
\caption{The longitudinal response function calculated in nuclear
matter for density $\rho=0.16$ fm$^{-3}$ and $|{\bold q}|=500$ MeV as a 
function of $\omega$. Dashed line: Hartree response with free dipole
form factors, dash-dotted line: Hartree response including the medium
corrections to the nucleon form factors, solid line: RPA response
including the medium corrections to the nucleon form factors. The NN
interaction used for the RPA calculation is the one shown in Fig.15,
but multiplied by meson-nucleon form factors including medium corrections
as explained in the main text.}
\end{center}
\end{figure}

The result for the longitudinal response function of NM 
($\rho=0.16$ fm$^{-3}$) for $|{\bold q}|=0.5$ GeV
is shown in Fig. 16. The dashed line is the Hartree response with the
free dipole form factors, and the dash-dotted line is obtained by
adding our calculated medium corrections $\Delta F(q^2;\rho) = F(q^2,\rho)
- F(q^2,\rho=0)$ to the free dipole form factors. (Here $F$ denotes any 
of the Dirac-Pauli form factors.) We see that, even in this region of
relatively low momentum transfers, the medium effects are appreciable.
Finally, the solid line shows the result obtained by further adding the
RPA corrections with the NN interaction derived above and using 
the density dependent meson-nucleon form factors. 

We do not show a comparison to experimental data in Fig.16 because
of two reasons: First, our calculation refers to NM and
cannot be applied directly to finite nuclei, although 
for the case of $^{40}$Ca the results are qualitatively very similar
to the NM results, see Ref.\cite{WAL}. Second, the analysis of the
experimental data is still controversial, mainly because of the model 
dependence of the Coulomb corrections\cite{JOUR,WILL,MM}
\footnote{The analysis of Ref.\cite{MEZ},
which did not take into account the Coulomb corrections, was refined 
in Ref.\cite{WILL} by using a particular theoretical model
for the Coulomb corrections, but it
has been pointed out recently \cite{MM} that the results depend on
which theoretical prescriptions are used.}.

\section{Summary}
In this paper we used a simple quark-scalar diquark 
picture for the single nucleon to describe the electromagnetic
form factors of a nucleon bound in the nuclear medium. 
We used the nuclear matter equation of state derived
within the same effective quark theory to assess the
effect of the mean nuclear fields on the internal quark
structure and the form factors of a bound nucleon, 
taking into account also the meson cloud around the 
constituent quarks. We have shown that this simple model 
gives reasonable results for the Dirac form factors
of the free nucleon if the finite
size of the diquark is taken into account. This is particularly
important for the neutron in order to obtain a small $F_{1}$,
consistent with observations. Concerning the Pauli form factors, 
in particular the anomalous magnetic moments and the behavior 
for large $Q^2$, one would need to further add the effect of the 
axial vector diquark and its finite size to achieve a reasonable 
description. 

The medium modifications of the form factors associated with the
orbital current are significant in the 
region of low and intermediate $Q^2$, and partially associated 
with an increase of the electric size 
in the medium. The form factors associated with the spin current
are enhanced because of the reduced nucleon mass, 
but due to a simultaneous decrease of the intrinsic anomalous magnetic moment, 
the total changes are not very large; in particular for the neutron 
they are almost zero. For both kinds of form factors - orbital and spin -, 
the medium modifications decrease with increasing $Q^2$. This is
consistent with the intuitive expectation that the mean
fields, which reflect the long range nuclear correlations, should not
influence the structure of the nucleon at short distances.

As an application, we considered the longitudinal response function of
nuclear matter for inclusive quasielastic electron scattering.
Even in this region of relatively low momentum transfer, the
medium effects on the nucleon form factors give rise to an
appreciable quenching of the response function. In order to
take into account the RPA-type correlations between the nucleons,
we derived the NN interaction in the present quark model for the
isoscalar channel, which is most important for the longitudinal
response function. In the limit of zero momentum transfer (Landau
limit), this interaction agrees with the more general Landau-Migdal
effective interaction, which was derived in earlier works for this
particular quark theory, and has many features which are similar
to relativistic meson-nucleon theories. In particular, it is
on the average repulsive in the region
below the quasielastic peak, and becomes attractive for higher
energy transfers.
            
We finally would like to remark that the language of effective
quark theories is very appropriate to address the problem of
medium modifications of nucleon properties. This work represents
one further step toward the goal of describing relativistic
nuclear systems by taking into account the quark substructure
of the constituent hadrons.

\vspace{0.7cm}

{\sc Acknowledgment} 

W.B. wishes to thank A.W. Thomas for many helpful discussions. 
This work was supported by the Grant in Aid for Scientific
Research of the Japanese Ministry of
Education, Culture, Sports, Science and Technology, Project No. C-16540267.

\newpage

\appendix
{\LARGE Appendices}

\section{Expressions for the nucleon form factors}
\setcounter{equation}{0}

Here we give the explicit expressions for the contributions of the
3 terms in Eq.(\ref{contact})-(\ref{diquark}) to the nucleon 
Dirac-Pauli form factors defined in
(\ref{par1}). They can be derived by invariant integration in the
usual way, i.e., by (i) introducing Feynman parameters, (ii) performing a
shift so that the denominator depends only on $k^2$, (iii) expressing the
result in the form (\ref{par1}) by using the Dirac equation for the
nucleon spinors, (iv) performing a Wick rotation according to
\begin{eqnarray}
-i \int \frac{{\rm d}^4 k}{(2\pi)^4} f(k^2) \longrightarrow
\int_0^{\infty} \frac{t {\rm d}t}{16 \pi^2} f(-t) \nonumber
\end{eqnarray} 
where $t=k_0^2+{\bold k}^2$ is the square of the Euclidean
length, 
and finally (v) introducing a Lorentz invariant regularization scheme, in our
case the proper time regularization (\ref{pt}). Below we give the
expressions which are obtained after performing the steps (i)-(iv),
using $Q^2=-q^2>0$ as the variable.

In the expressions given below there enter the diquark and nucleon
wave function normalization factors, which are defined by
(\ref{gs}) and (\ref{zn}) in terms of the renormalized bubble
graphs (\ref{newbubbs}) and (\ref{newbubbn}), for which we have the
expressions
\begin{eqnarray}
\hat{\Pi}_s(p^2) &=& - 24 \int_0^{\infty} \frac{t {\rm d}t}{16 \pi^2}
\left( \frac{1}{t+M^2} + \frac{p^2}{2} \int_0^1 {\rm d}x 
\frac{1}{\left(t+M^2-p^2 x(1-x)\right)^2} \right) \nonumber \\
\label{pisc} \\
\hat{\Pi}_N(p) &=& - \hat{g}_s \int_0^{\infty} \frac{t {\rm d}t}{16 \pi^2}
\int_0^1 {\rm d}x \frac{\fslash{p} x + M}
{\left(t + M^2(1-x) + M_s^2 x - p^2 x(1-x)\right)^2}\,. \nonumber \\
\label{pinuc} 
\end{eqnarray} 

If we denote
\begin{eqnarray}
D_1(Q^2,x) = t + M^2 + Q^2 x(1-x)\,, \label{d1}
\end{eqnarray} 
the contributions of the contact term (\ref{contact}) to the form factors
are as follows:
\begin{eqnarray}
F_{1N}^{(C)}(Q^2) &=& - \frac{4 G_s}{\hat{g}_s} \hat{Z}_N 
\int_0^{\infty} \frac{t {\rm d}t}{16 \pi^2} \int_0^1 {\rm d}x 
\nonumber \\
&\times&  
\left[F_{1Q}(Q^2) \left( \frac{1}{D_1(Q^2,x)} - \frac{1}{t+M^2}
+ \frac{Q^2}{2}\, \frac{1}{D_1(Q^2,x)^2} \right) \right. 
\nonumber \\
&+& \left. F_{2Q}(Q^2)\, \frac{Q^2}{2}\, \frac{(1+x)(1-x\frac{M_N}{M})}
{D_1(Q^2,x)^2} \right] \label{f1c} \\
F_{2N}^{(C)}(Q^2) &=&  \frac{4 G_s}{\hat{g}_s} \hat{Z}_N  \, M_N M
\int_0^{\infty} \frac{t {\rm d}t}{16 \pi^2} \int_0^1 {\rm d}x 
\nonumber \\ 
&\times& \left[F_{1Q}(Q^2) \, \frac{2}{D_1(Q^2,x)^2} 
+ F_{2Q}(Q^2)\,  \frac{(1-x\frac{M_N}{M})^2 - 
\frac{Q^2x}{M^2}}{D_1(Q^2,x)^2} \right].\nonumber \\
\label{f2c}
\end{eqnarray}
    
If we denote
\begin{eqnarray}
D_2(Q^2,x,y) &=& t + M_s^2(1-x) + M^2 x - M_N^2 x(1-x) +
\frac{Q^2}{4}(x^2-y^2)  \nonumber \\
N_1(Q^2,x) &=& (M_N+M)^2 - M_s^2 - \frac{t}{2} - 2 M_N^2
x(1-x) - 2 M_N M x \nonumber \\
N_2(Q^2,x) &=& \left(1+\frac{M_N}{M}(1-x)\right)^2 - \frac{Q^2}{4M^2}(x^2-y^2)
\,, \nonumber \\
\label{d2}
\end{eqnarray}
the contributions of the quark term (\ref{quark}) to the form factors
are as follows:
\begin{eqnarray}
F_{1N}^{(Q)} &=& \hat{Z}_N \int_0^{\infty} \frac{t {\rm d}t}{16 \pi^2} 
\int_0^1 {\rm d}x  \nonumber \\
&\times& 
\left[ F_{1Q}(Q^2) \left(  \frac{1}{D_1(Q^2,x)^2} + \int_{-x}^{x} {\rm d}y 
\frac{N_1(Q^2,x)}{D_2(Q^2,x,y)^3} \right) \right. \nonumber \\
&-& \left. F_{2Q}(Q^2) \frac{Q^2}{2} \int_{-x}^{x} {\rm d}y \, 
\frac{1+\frac{M_N}{M} (1-x)}{D_2(Q^2,x,y)^3} \right] \label{f1q} \\
F_{2N}^{(Q)} &=& \hat{Z}_N \int_0^{\infty} \frac{t {\rm d}t}{16 \pi^2} 
\int_0^1 {\rm d}x \int_{-x}^{x} {\rm d}y  \nonumber \\
&\times& \left[ F_{1Q}(Q^2)\, 2M_N x \, \frac{M_N(1-x)+M}{D_2(Q^2,x,y)^3} 
+ F_{2Q}(Q^2) \, M_N M \, \frac{N_2(Q^2,x)}{D_2(Q^2,x,y)^3} \right]\,. 
\nonumber \\
\label{f2q}
\end{eqnarray}

If we denote
\begin{eqnarray}
D_3(Q^2,x,y) = t + M^2(1-x) + M_s^2 x  - M_N^2 x(1-x) +
\frac{Q^2}{4}(x^2-y^2)\,, \nonumber \\
\label{d3}
\end{eqnarray}
the contributions of the diquark term (\ref{diquark}) to the form factors
are as follows:
\begin{eqnarray}
F_{1N}^{(D)} &=& \hat{Z}_N F_D(Q^2) \int_0^{\infty} 
\frac{t {\rm d}t}{16 \pi^2} \int_0^1 {\rm d}x \int_{-x}^{x} {\rm d}y
\frac{2 M_N(1-x)(M_Nx+M)+\frac{t}{2}}{D_3(Q^2,x,y)^3} \nonumber \\
\label{f1d} \\
F_{2N}^{(D)} &=& - \hat{Z}_N F_D(Q^2) \int_0^{\infty} 
\frac{t {\rm d}t}{16 \pi^2} \int_0^1 {\rm d}x \int_{-x}^{x} {\rm d}y
\, 2 M_N (1-x)\,  \frac{M_N x +M}{D_3(Q^2,x,y)^3}\,. \nonumber \\
\label{f2d}
\end{eqnarray}
The diquark form factor $F_D$ is calculated from the expression (\ref{vert})
as
\begin{eqnarray}
F_D &=& 12 \hat{g}_s  \int_0^{\infty} 
\frac{t {\rm d}t}{16 \pi^2} \int_0^1 {\rm d}x
\left\{ F_{1Q}^{(0)}(Q^2) \left[ \frac{1}{D_1(Q^2,x)^2} +
M_s^2 \int_{-x}^{x} {\rm d}y \frac{x}{D_4(Q^2,x,y)^3} \right] \right. 
\nonumber \\
&-& \left. \frac{Q^2}{2} F_{2Q}^{(0)}(Q^2) \int_{-x}^{x} {\rm d}y 
\frac{1}{D_4(Q^2,x,y)^3} \right\}  \label{fd}
\end{eqnarray}
where we defined
\begin{eqnarray}
D_4(Q^2,x,y) = t + M^2 - M_s^2 x(1-x) + \frac{Q^2}{4} (x^2-y^2)\,, 
\label{d4}
\end{eqnarray}
and $F_{1Q}^{(0)}$, $F_{2Q}^{(0)}$ are the isoscalar parts of the
quark form factors.  

\section{Expressions for the quark form factors}
\setcounter{equation}{0}
Here we give the expressions for the quark electromagnetic vertex
(\ref{jq1}) in terms of the quark form factors defined in Eq.(\ref{quarkff}). 
In the expressions given below there enter the pion and quark wave
function normalizations $\hat{g}_{\pi}$ and $\hat{Z}_Q$, which are
defined in terms of the renormalized self energies $\hat{\Pi}_{\pi}$ 
and $\hat{\Sigma}_Q$ by Eqs. (\ref{gpi}) and (\ref{zq2}). The
expression for $\hat{\Pi}_{\pi}=\hat{\Pi}_{s}$ has been given in
Eq.(\ref{pisc}), and 
\begin{eqnarray}
\hat{\Sigma}_Q(p) &=& - 3 \hat{g}_{\pi} \int_0^{\infty} 
\frac{t {\rm d}t}{16 \pi^2}
\int_0^1 {\rm d}x \frac{\fslash{p} x - M}
{\left(t + M^2(1-x) + M_{\pi}^2 x - p^2 x(1-x)\right)^2} \nonumber \\
\label{piquark} 
\end{eqnarray}
As explained in the main text, the isoscalar (or isovector) parts of the 
quark vertices given below should eventually be further multiplied by the
VMD form factors (\ref{vmd}), where the expression for $\hat{\Pi}_V$ is
\begin{eqnarray}
\hat{\Pi}_V(Q^2) = 48 \, Q^2 \int_0^{\infty} 
\frac{t {\rm d}t}{16 \pi^2} \int_0^1 {\rm d}x \, 
\frac{x(1-x)}{D_1(Q^2,x)^2}\,.
\label{piv}
\end{eqnarray}
  
The quark diagram (second term in (\ref{jq1})) gives the following 
contributions to the quark form factors:
\begin{eqnarray}
F_{1Q}^{(Q)} &=& \frac{1}{2}\left(1-\tau_3\right) \hat{g}_{\pi}
\int_0^{\infty} \frac{t {\rm d}t}{16 \pi^2} \int_0^1 {\rm d}x 
\left[\frac{1}{D_1(Q^2,x)^2} + \int_{-x}^{x} {\rm d}y 
\frac{2 M^2 x^2 - M_{\pi}^2 - \frac{t}{2}}{D_5(Q^2,x,y)^3} \right]
\nonumber \\
\label{f1qq} \\
F_{2Q}^{(Q)} &=& -\frac{1}{2}\left(1-\tau_3\right) \hat{g}_{\pi}
\int_0^{\infty} \frac{t {\rm d}t}{16 \pi^2} 
\int_0^1 {\rm d}x \int_{-x}^{x} {\rm d}y \, 
\frac{2 M^2 x^2}{D_5(Q^2,x,y)^3}
\label{f2qq} 
\end{eqnarray}
where
\begin{eqnarray}
D_5(Q^2,x,y) = t + M_{\pi}^2(1-x) + M^2 x^2 + \frac{Q^2}{4}(x^2-y^2)\,. 
\label{d5}
\end{eqnarray}

The pion diagram (third term in (\ref{jq1})) gives the following contributions
to the quark form factors:
\begin{eqnarray}
F_{1Q}^{(\pi)} &=& -2 \tau_3  \, \hat{g}_{\pi}\, F_{\pi}(q^2) 
\int_0^{\infty} \frac{t {\rm d}t}{16 \pi^2} 
\int_0^1 {\rm d}x \int_{-x}^{x} {\rm d}y \,
\frac{2 M^2 (1-x)^2 - \frac{t}{2}}{D_6(Q^2,x,y)^3}
\nonumber \\
\label{f1qpi} \\
F_{2Q}^{(\pi)} &=& 4\, \tau_3 \, M^2 \, \hat{g}_{\pi} \, F_{\pi}(q^2) 
\int_0^{\infty} \frac{t {\rm d}t}{16 \pi^2} 
\int_0^1 {\rm d}x \int_{-x}^{x} {\rm d}y \,
\frac{(1-x)^2}{D_6(Q^2,x,y)^3} \,, \nonumber \\
\label{f2qpi}
\end{eqnarray}
where 
\begin{eqnarray}
D_6(Q^2,x,y) = t + M^2(1-x)^2 + M_{\pi}^2x + \frac{Q^2}{4}(x^2-y^2)\,. 
\label{d6}
\end{eqnarray}
The pion form factor is calculated from (\ref{pionff}) and (\ref{lpi}) as
\begin{eqnarray}
F_{\pi} &=& 12 \hat{g}_{\pi} \int_0^{\infty} 
\frac{t {\rm d}t}{16 \pi^2} \int_0^1 {\rm d}x
\left[ \frac{1}{D_1(Q^2,x)^2} +
M_{\pi}^2 \int_{-x}^{x} {\rm d}y \frac{x}{D_7(Q^2,x,y)^3} \right]\,, 
\nonumber \\
\label{piff}
\end{eqnarray}
where
\begin{eqnarray}
D_7(Q^2,x,y) = t + M^2 - M_{\pi}^2 x(1-x) + \frac{Q^2}{4}(x^2-y^2) \,.
\label{d7}
\end{eqnarray}

\section{The NN interaction}
\setcounter{equation}{0}
Here we provide some details on the NN interaction Eqs.(\ref{vnn}).

The bubble graph in the sigma channel, which appears in (\ref{sigma}),
has the form
\begin{eqnarray}
\hat{\Pi}_{\sigma}(k^2) &=& 6i \int \frac{{\rm d}^4q}{(2\pi)^4} {\rm Tr}_D 
\left(S_F(q) S_F(k+q) \right) \nonumber \\
&=& 12 \int_0^{\infty} \frac{t {\rm d}t}{16 \pi^2} \left[
\frac{2}{t+M^2} + \left(k^2-4M^2\right) \int_0^1 {\rm d}x
\frac{1}{\left(t+M^2 - k^2x(1-x)\right)^2} \right]\,. 
\nonumber \\
\label{pis}
\end{eqnarray}
The expression for $\hat{\Pi}_V$ has been given in (\ref{piv}). 

In the calculations of Sect.5, the NN interaction (\ref{vnn}) is used
without further approximations. Here we wish to give an approximate form
in terms of Yukawa potentials: Expanding
$\hat{\Pi}_{\sigma}$ and $\hat{\Pi}_V$ around $k^2=0$, we obtain
\begin{eqnarray}
V_{\rm NN}(k) = \frac{g_{\sigma N}}{k^2-M_{\sigma}^2} 
({\bold 1}) \cdot ({\bold 1}) - \frac{g_{\omega N}}{k^2-M_{\omega}^2} 
({\gamma_{\mu}}) \cdot ({\gamma^{\mu}})\,, \label{v}
\end{eqnarray}
where
\begin{eqnarray}
g_{\sigma N} &=& \left(\frac{{\rm d}M_N}{{\rm d}M}\right)^2 g_{\sigma}
= (2.34)^2 \times 21.75 = 119.4 \label{gsigma} \\
g_{\omega N} &=& 9 \, g_{\omega}
= 9 \times 18.43 = 165.9 \,, \label{gomega}
\end{eqnarray}
where the quark-meson couplings are defined by \\
$g_{\sigma}=\left[\left(\partial \hat{\Pi}_{\sigma}/\partial k^2\right)
_{k^2=0}\right]^{-1}$ and 
$g_{\omega}=\left[- \left(\partial \hat{\Pi}_{V}/\partial k^2\right)_{k^2=0} 
\right]^{-1}$, and the numerical values given above are obtained from
our nuclear matter EOS for $\rho=0.16$ fm$^{-3}$. The meson masses 
defined at zero momentum are
\begin{eqnarray}
M_{\sigma}^2 &=& g_{\sigma} \left(\frac{1}{2 G_{\pi}} - \hat{\Pi}_{\sigma}(0)
+ \delta M_{\sigma}^2 \right) = \left(0.81\,{\rm GeV}\right)^2 \label{msi} \\
M_{\omega}^2 &=& \frac{g_{\omega}}{2 G_{\omega}} = 
\left(1.13\,{\rm GeV}\right)^2\,. \label{mom}
\end{eqnarray}
Note that these masses are different from the pole positions. 
 
\end{document}